# Effects of Long-chain Branching, Short-chain Branching, and Polydispersity on Pressure Sensitive Rheology of Polymer Melts

Lilian Lin, Matthew Joe, and Heon E. Park (박희언)*

Department of Chemical and Paper Engineering, Western Michigan University, 4601 Campus Dr., Kalamazoo, MI 49008 USA
* Corresponding author: heon.park@wmich.edu

**Abstract**
The rheological behavior of polymer melts under high pressure is a critical factor in many industrial processes like injection molding and extrusion, yet it is often inadequately characterized. At operating pressures that can exceed 100 MPa, viscosity can increase by orders of magnitude, making atmospheric-pressure data insufficient for accurate process simulation. This pressure-induced viscosity increase is highly dependent on molecular architectures of the materials. This study aims to deconstruct the influence of specific structural features such as short-chain branching (SCB), long-chain branching (LCB), and polydispersity (PDI) on the pressure sensitivity of the viscosity of polyethylene. Utilizing a high-pressure sliding plate rheometer (HPSPR) to ensure accurate measurements under uniform shear and pressure, we characterized four distinct polyethylene melts. All samples, regardless of their structure, exhibited piezorheologically simple behavior, allowing the application of time–pressure superposition over the entire shear rate range. A key finding is that the long-chain branched sample, known from the literature to be thermorheologically complex, was found to be piezorheologically simple. This dichotomy is explained by the different physical mechanisms of temperature and pressure. The pressure sensitivity of the viscosity, quantified by the pressure–viscosity coefficient, was found to be strongly dependent on molecular branching. Both SCB and LCB significantly increase the pressure sensitivity while polydispersity had a negligible effect. These results demonstrate that molecular branches are the dominant structural parameter controlling the rheological response of polyethylene to pressure, providing crucial insights for the development of more accurate predictive models for high-pressure polymer processing.

## 1. Introduction

A central goal of current research is to establish clear structure–property relationships. However, material properties cannot be adequately described without specifying their state or the operating conditions where those are measured. Thus, a more complete paradigm is structure–processing–property, where rheology plays a key role [1-3] along with mechanical strengths [4-7]. Pressure is a fundamental variable that significantly influences the physics and chemistry of materials. It affects a wide range of physical and chemical phenomena, including capillary flow [8], micro/nano fluidics [9,10], cavitation [11], crystallization [12], and synthesis [13]. In polymer science, pressure, along with temperature, profoundly alters the melt and solution rheology [14-17]. This effect is especially critical in plastics manufacturing. In processes like high-speed injection molding, pressures can reach 100 MPa, causing the polymer melt viscosity to increase by an order of magnitude or more. Consequently, viscosity data obtained at atmospheric pressure can be highly misleading when applied to high-pressure processes. Accurate high-pressure rheological data are essential for the simulation and design of injection molds and are also necessary for understanding other processes like micro injection molding [18] and wall slip [19]. The need for high-pressure rheology extends to other materials as well, such as gels [20-23], particulate composites [24], materials with supercritical fluids [14,25,26], and molecular assemblies [27]. Despite the clear need for accurate data, obtaining reliable high-pressure rheological measurements is challenging [20,28]. While there have been efforts in theoretical modeling [29,30] and using AI for rheological prediction [31], experimental data remains sparse. Various experimental methods exist [28,32-36], but many have significant limitations. For instance, flow and pressure are often not homogeneous in capillary rheometers [37], leading to inconsistent results between different instruments [38]. Rotational rheometers designed for high-pressure work also have their own set of limitations and restrictions [20,25]. On the other hand, the response of a polymer to operating conditions like pressure and temperature is also heavily influenced by its molecular structure, including its polydispersity, short-chain branching (SCB), and long-chain branching (LCB). In this study, we utilize a unique scientific instrument, the high-pressure sliding plate rheometer (HPSPR), to investigate the effect of pressure on the viscosity of polymer melts. By selecting polymers with varied molecular structures, we aim to understand how these structural differences affect the pressure sensitivity of viscosity.

### 1.1. Pressure–Volume–Temperature Behavior

Prior to rheological measurements, the pressure–volume–temperature behavior (PVT) of each sample was determined at various temperatures and pressures, mainly for two reasons. Firstly, it should be confirmed that each sample is in a molten state at elevated pressures and especially at the operating temperature. These high-pressure measurements must be carried out at temperatures well above the melting point ($T_m$) to avoid the pressure-induced crystallization that can occur as $T_m$ increases with pressure. For this study, the pressure dependencies of $T_m$ were obtained by determining the PVT behavior of the polymer. Secondly, the specific volume at the operating temperature and pressures is necessary to obtain the vertical shift factor [1,14]. The experiments were performed isothermally while increasing pressure, with Wohl's modified version [14,39] of the empirical Tait equation [40] being used to describe the PVT behavior:

$$v(P,T) = v(0,T)\left\{1 - C(T)\ln\left[1 + \frac{P}{B(T)}\right]\right\} \quad (1)$$

where the zero-pressure isotherm used in this research is given by:

$$v(0,T) = v_0 + v_1 T \quad (2)$$

and $B(T)$ is given by:

$$B(T) = B_0 e^{-B_1 T} \quad (3)$$

where $v_0$, $v_1$, $B_0$, and $B_1$ are fitting constants, with $C(T) = 0.0894$ [41]. The prediction of the modified Tait model is required for this study since the rheological data were measured at pressures different from the PVT measurement pressures, which were set by the operating software. To obtain the vertical shift factors $b_{T,P}(T,P)$ at various temperatures and pressures, the density of the polymers at the conditions, where rheological data are measured, is necessary:

$$b_{T,P}(T,P) \equiv \frac{\rho(T,P)T}{\rho(T_0,P_0)T_0} \quad (4)$$

where $T_0$ and $P_0$ are the reference temperature and pressure, respectively. Since the rheological behavior is determined at only one temperature in this study, the vertical shift factor that was used is:

$$b_P(P) \equiv \frac{\rho(P)}{\rho(P_0)} = \frac{v(P_0)}{v(P)} \quad (5)$$

In standard fluid mechanics, liquids (and gases at Ma < 0.3) are generally treated as incompressible because the effect of pressure on their density is negligible. For polymer melts, however, this assumption is invalid, as pressure-induced density changes have a significant effect on their rheology. This context necessitates a clear distinction regarding the term "pressure." The pressure, $P$, in this study is not the gauge-free "hydrodynamic pressure" that appears in the incompressible Navier-Stokes equations. In that context, hydrodynamic pressure acts as a Lagrange multiplier to enforce incompressibility ($\nabla \cdot \boldsymbol{v} = 0$), and only its gradient enters the momentum balance [42]. Rather**,** the pressure discussed here is the absolute, spatially uniform hydrostatic (thermodynamic) pressure**.** This is the pressure externally imposed on the specimen within a sealed high-pressure vessel, such as a PVT cell or a the HPSPR, where no pressure gradient exists. In these instruments, the sample is confined and pressurized by an inert fluid (e.g., mercury or inert oil) until a prescribed setpoint pressure is reached. This hydrostatic pressure is the critical variable that sets the material's specific volume and modifies its viscoelastic relaxation times. Although molten polyethylenes are sometimes treated as incompressible in macroscopic flow calculations, their bulk modulus is finite, and their specific volume decreases measurably with pressure. For instance, standard PVT data for polyolefins show volume changes of several percent up to 100–200 MPa [43], which is consistent with the 6% density change we observe up to 70 MPa (Figure 1). Our study is therefore concerned with this absolute hydrostatic pressure, as captured by the Tait equation (for specific volume) and pressure-dependent shift factors (for relaxation times).

### 1.2. Shear Thinning and Effect of Polydispersity

Typically, polymer melts exhibit shear-thinning behavior, where the viscosity falls with shear rate. This tends to occur as polymer chains align in the direction of flow, reducing entanglements and lowering viscosity. During the steady simple shear tests for polymer melts, the stress growth function with time, i.e., $\sigma^{+}(t, \dot{\gamma}, P)$, does not reach its steady state value, $\sigma(\dot{\gamma}, P)$ instantaneously due to the viscoelastic behavior of the polymer. However, once $\sigma(\dot{\gamma}, P)$ was obtained, and this value can be used to calculate the viscosity:

$$\eta(\dot{\gamma}, P) = \frac{\sigma(\dot{\gamma}, P)}{|\dot{\gamma}|} \quad (6)$$

For a fluid in steady shear, the shear stress is an odd function of the shear rate, and thus using the magnitude of shear rate can guarantee that the viscosity depends only on the magnitude of the shear rate. There are a number of models relating the viscosity and shear rate (or the magnitude of complex viscosity and frequency). In this study, once the stress versus shear rate at $P_0 = 0.1$ MPa was obtained, the following Cross model for stress [44] was used to fit the data:

$$\sigma(\dot{\gamma}, P_0) = \frac{\eta_0(P_0)\dot{\gamma}}{1 + |\lambda(P_0)\dot{\gamma}|^{m(P_0)}} \quad (7)$$

while the viscosity versus shear rate was obtained by the Cross model for viscosity:

$$\eta(\dot{\gamma}, P_0) = \frac{\eta_0(P_0)}{1 + |\lambda(P_0)\dot{\gamma}|^{m(P_0)}} \quad (8)$$

where $\eta_0$ is the zero-shear viscosity in Pa s, $\lambda$ is the characteristic time in s, and $m$ is the fitting constant. These three parameters were determined by the least square method.

Polydispersity (PDI), the characteristic variable representing the breadth of the molecular weight (molar mass) distribution (MWD), significantly influences the rheological behavior of polymer melts, particularly shear thinning [45-47], where a higher PDI indicates a wider range of chain lengths, which affects chain entanglement and relaxation dynamics. PDI is defined as:

$$\mathrm{PDI} \equiv \frac{M_\mathrm{w}}{M_\mathrm{n}} \quad (9)$$

where $M_\mathrm{w}$ and $M_\mathrm{n}$ are weight-average and number-average molecular weights, respectively. In this study, polydispersity denotes the MWD of the whole molecules, which are considered with their backbone and any branches as a single entity. As polydispersity increases, shear thinning increases by enabling shorter chains to function as molecular lubricants, thereby facilitating the disentanglement and flow of longer, entangled chains under applied shear. This results in an earlier (i.e., at low shear rates) onset of shear thinning and a more pronounced reduction in viscosity as shear rate increases, compared to monodisperse polymers. Flow curves for polydisperse melts typically exhibit a Newtonian plateau at low shear rates, followed by a shear-thinning region, and power-law region at high shear rates. The zero-shear viscosity increases with PDI when maintaining a constant $M_\mathrm{n}$ as the presence of longer chains contributes to greater entanglement density. However, at elevated shear rates, higher PDI promotes more significant conformational changes in longer chains, leading to lower viscosities. As PDI increases, the Newtonian plateau region shortens, and the shear-thinning behavior becomes more evident, as predicted by superposition models relating MWD to viscosity.

### 1.3. Effect of Pressure on Rheological Properties

A central goal in modern physics, chemistry, and materials processing is to connect the structure of a material to its properties. However, since processing conditions profoundly affect the state of the material, it is equally critical to understand how these conditions, in conjunction with structure, dictate the final properties. Among these variables, pressure plays a crucial role, with the practical importance of pressure-dependent viscosity spanning multiple fields: for example, in polymer processing, operations like injection molding and extrusion rely on the controlled flow of the molten polymer and its resistance to flow is quantified by viscosity. Thus, accurate prediction of pressure-dependent viscosity behavior is essential for process simulation, optimization, and quality control. However, this task is challenging due to its high sensitivity to external conditions, particularly high temperature and pressure. Hence, high-pressure rheometry is essential for generating the accurate data needed for process simulations, where key data includes the pressure–viscosity coefficient, flow curves, and shear-thinning behavior under high pressure.

Two major theoretical frameworks describe the effect of pressure on viscosity and viscoelastic properties [48]. Firstly, the absolute rate/transition state theory was originally developed to describe the effect of temperature on chemical reaction rates [49]. This approach was then extended to model the physical processes of flow [50], and later to incorporate pressure [51]. In this theory, both temperature and pressure alter the response time of molecular segments to external force, where increasing pressure or lowering temperature slows down segmental rearrangements, thereby exponentially increasing viscosity. At the molecular level, pressure raises the activation energy barrier, making it harder for molecules to overcome intermolecular interactions and move past their neighbors. The empirical Barus model compliments this theoretical framework [52], describing the exponential increase of viscosity with pressure:

$$\eta(P) = Be^{\beta P} \qquad 10$$

where $B$ is a pre-exponetial coefficient, $\beta$ is the pressure coefficient, and $P$ is again the spatially uniform hydrostatic pressure in the sample. Because only pressure differences appear in the Barus form, any arbitrary additive constant in pressure (as in incompressible hydrodynamics) is immaterial. What matters for the rheology is the change in hydrostatic pressure relative to the reference state, not relative to another location, because it alters specific volume and the activation volume for segmental rearrangements.

Secondly, the free volume theory was developed to relate viscosity to the available free volume in a material [53]. This concept was then used to model the effect of temperature on viscosity via the Williams–Landel–Ferry (WLF) equation [15,54,55]. Subsequently, the theory was extended to account for the effect of pressure, leading to the Ferry–Stratton model [56,57]:

$$\ln \eta(P) = A' + \frac{B_0 v_0 f(P)}{v_P - v_0 f(P)} \qquad 11$$

where $A'$ and $B_0$ are temperature-invariant parameters related to the characteristics of molecules, $v_P$ is the specific volume at $P$, $v_0$ is the specific volume occupied by molecules, i.e., limiting specific volume extrapolated to 0 K, and $f(P)$ is a quadratic function of pressure, accounting for the effect of pressure on the free volume. In this approach, temperature and pressure govern flow by modifying free volume. This model is especially useful near the glass transition temperature or at very high pressures, where free volume is scarce and flow is governed by the availability of "holes." While the free volume theory predicts that a polymer's viscosity should be constant for any combination of temperature and pressure that yields the same specific volume, this is often not observed experimentally. A polymer can exhibit different viscosities at two distinct conditions even when the specific volumes are identical. This demonstrates that the relationship between viscosity and free volume is not a one-to-one correspondence and suggests that viscosity is influenced by factors other than free volume alone.

It is instructive to compare the effects of pressure and temperature on material properties [1,2,21,55,58,59]. Both pressure and temperature fundamentally affect viscosity [60] by altering the free volume or/and segmental mobility of polymer chains. Phenomenologically, decreasing temperature has a similar effect as increasing pressure, but they do so through different mechanisms and with different sensitivities. For many polymers within typical processing windows, the sensitivity of viscosity to pressure is comparable to, and sometimes greater than, its sensitivity to temperature [61].

For example, a pressure increase of 100 MPa can have a similar effect on viscosity as a temperature decrease of approximately 40°C [62,63]. In typical injection molding, pressure can vary from near 0 to over 150 MPa, while temperature is often controlled within a 30°C window. This highlights that pressure effects can be as significant as, or even more significant than, temperature effects Time–Pressure Superposition.

The effect of pressure on the rheology can be quantitatively expressed by pressure shift factors based on time–pressure superposition. This principle states that increasing pressure has the same effect on the relaxation spectrum of a polymer as decreasing temperature or decreasing the frequency of deformation. Viscosity or viscoelastic data (such as storage and loss moduli) measured at various pressures (or temperatures) can be shifted along the stress axis and along the shear rate or frequency axis to form a single master curve which is normally based on reference pressure (or temperature). The factors shifting the high-pressure data to the reference pressure are known as the vertical and horizontal pressure shift factors, respectively. To perform this, first, reduced stress is obtained utilizing the vertical shift factors (Eq. 5) as:

$$\sigma_r(\dot{\gamma}, P) \equiv \frac{\sigma(\dot{\gamma}, P)}{b_P(P)} \qquad 12$$

which naturally shifts the stress data at high pressures vertically (i.e. in stress axis). Second, the horizontal shift (i.e. in shear rate axis) of $\sigma_r(\dot{\gamma}, P)$ was performed by determining $a_P(P)$ to minimize the summation of the squared differences between the horizontally shifted data and the Cross model predictions for stress:

$$\sigma_r(\dot{\gamma}, P) = \frac{\sigma(\dot{\gamma}, P)}{b_P(P)} = \frac{\eta_0(P_0)a_P(P)|\dot{\gamma}|}{1 + |\lambda(P_0)a_P(P)\dot{\gamma}|^{m(P_0)}} \qquad 13$$

where fitting parameters, $\eta_0(P_0), \lambda(P_0)$, and $m(P_0)$, are independent of pressure, and thus parameters obtained for 0.1 MPa are valid for data at the other pressures. To obtain $a_P(P)$, the reduced stress (not viscosity) versus shear rate data were used since only a shift in the shear rate axis is necessary while viscosity versus shear rate data require shifts in both viscosity and shear rate axes. The zero-shear viscosity at each pressure can be obtained by:

$$\eta_0(P) = \eta_0(P_0)a_P(P)b_P(P) \qquad 14$$

and the characteristic time at each pressure by:

$$\lambda(P) = \lambda(P_0)a_P(P) \qquad 15$$

Equation 13 represents the model equation for the master curve of stress at 0.1 MPa, and the model equation for the master curve of the viscosity based on the viscosity at various pressures can be easily obtained by:

$$\frac{\eta(\dot{\gamma}, P)}{a_P(P)b_P(P)} = \frac{\eta_0(P_0)}{1 + |\lambda(P_0)a_P(P)\dot{\gamma}|^{m(P_0)}} \qquad 16$$

with $a_P(P)$ being used to shift the viscosity versus shear rate, $\eta(\dot{\gamma}, P)/b_P(P)$ in both axes. The model equation for the master curve for stress based on the data at various pressures becomes:

$$\frac{\sigma(\dot{\gamma}, P)}{a_P(P)b_P(P)} = \frac{\eta_0(P_0)|\dot{\gamma}|}{1 + |\lambda(P_0)a_P(P)\dot{\gamma}|^{m(P_0)}} \qquad 17$$

Thus, we can plot the master curves at the reference pressure by plotting $\frac{\eta(\dot{\gamma},P)}{a_P(P)b_P(P)}$ or $\frac{\sigma(\dot{\gamma},P)}{a_P(P)b_P(P)}$ versus $a_P(P)\dot{\gamma}$. On the other hand, Eq. 18 shows predicted viscosity at high pressure as a function of shear rate, and this can be shown by plotting $\eta(\dot{\gamma}, P)$, obtained by Eq. 18, versus $\dot{\gamma}$:

$$\eta(\dot{\gamma}, P) = \frac{\eta_0(P_0)a_P(P)b_P(P)}{1 + |\lambda(P_0)a_P(P)\dot{\gamma}|^{m(P_0)}} \qquad 18$$

The horizontal pressure shift factor can naturally be obtained by data and Eq. 14 as:

$$a(P) = \frac{\eta_0(P)}{\eta_0(P_0)b_P(P)} \qquad 19$$

Since the operating temperature is well above the melting temperature, an exponential equation that is equivalent to Barus' relationship [38,52] can be used to describe the pressure shift factor as a function of pressure:

$$\ln\left[\frac{\eta_0(P)}{\eta_0(P_0)}\right] = \ln[a_P(P)b_P(P)] = \beta(P - P_0) \qquad 20$$

where $\beta$ is the pressure coefficient of the viscosity defined as:

$$\beta \equiv \frac{1}{\eta_0}\left[\frac{\mathrm{d}\eta_0(P)}{\mathrm{d}P}\right] \qquad 21$$

Equation 20 is useful when the temperature is substantially higher than the glass transition temperature ($T_g$), which allows for enough space for polymeric chains to move.

### 1.4. Piezorheological and Thermorheological Behavior

For many linear homopolymers, viscosity data at different pressures can be shifted to a common reference (e.g., 0.1 MPa) using a single shift factor, $a_P(P)$ across the shear-rate range, demonstrating piezorheologically simple behavior [64,65]. However, piezorheological complexity arises when time–pressure superposition (TPS) fails, typically in heterogeneous polymers such as blends [66], block copolymers [58], layered composites [67,68], and hybrid materials [69,70]. The analogy extends to thermorheological complexity, where time–temperature superposition (TTS) fails, i.e., linear viscoelastic or shear data at various temperatures cannot be shifted to a reference temperature by a single temperature shift factor, $a_T(T)$ across the frequency or shear-rate range. Like piezorheologically complex materials, multi-phase materials often show thermorheologically complex behavior. However, even neat linear polymers may exhibit thermorheological complexity, especially when relaxation is dominated by the glassy modulus at low temperatures or high frequencies [71,72]. Thermorheologically complex behavior at relatively high temperatures or low frequencies can be also seen for long-chain branched polymers [73,74]. It is of interest to determine if the same material can exhibit thermorheological complexity but piezorheological simplicity, or *vice versa*. This study investigates this possibility.

### 1.5. Measurement Methods

#### 1.5.1. Challenges in High-Pressure Rheometry

Various experimental methods exist for conducting rheometry at high pressure [28,32-36]. However, performing rheological measurements at elevated pressures presents several fundamental and practical challenges [32,34,75]. Pressure generation and containment necessitate specialized pressure vessels and sealing mechanisms to ensure operator safety while minimizing compliance and distortion of the measurement geometry. Temperature control is equally critical if gradients within the pressurized sample chamber can induce property variations. Accurate transducer design is essential. Torque, force, and displacement sensors must endure high static loads while maintaining sensitivity to subtle viscoelastic responses. Furthermore, sample loading and handling pose difficulties as trapped air bubbles or incomplete filling can introduce artifacts in both steady and oscillatory measurements.

#### 1.5.2. High-Pressure Rotational and Plate Rheometry

Rotational rheometers equipped with high-pressure cells, utilizing cone-plate or parallel-plate geometries within a sealed pressure vessel, enable steady and oscillatory measurements under pressure. The strengths of this method include its flexibility, allowing for oscillatory, creep, or steady-shear tests, and its capability to measure viscoelastic spectra under controlled pressure. However, the pressure cells diminish torque sensitivity, thereby limiting the resolution of low-shear or small-amplitude oscillatory measurements [20,25,28,75]. Moreover, sealing complexity and the requirement for precise temperature control within the pressure cell represent additional limitations. Upon pressurization, the sample volume will shrink, creating a gap between the upper and lower plates and thereby leading to a reduction in measured stress. If a gas is employed as the pressurizing medium, it may dissolve into the edges of the sample. Given that the highest torque is generated at the sample edge, these scenarios can result in a significant underestimation of stress.

#### 1.5.3. High-Pressure Capillary Rheometry

Capillary rheometers with integrated back-pressure chambers are commonly used to determine pressure-dependent viscosity under conditions that simulate polymer processing [76,77]. These instruments assess the volumetric flow rate under a controlled pressure drop across a die, providing access to shear rates within industrially relevant ranges. The method is robust but demands careful corrections. Failure to apply these corrections results in systematic errors, especially at high pressures where end effects are amplified. Additionally, compressibility and viscous heating must be accounted for to prevent bias in viscosity estimation. Another limitation is that the shear rate, temperature, pressure, and shear stress are not uniform throughout the sample in the capillary [37]. Furthermore, different types of capillary rheometers may yield varying results due to design differences [38].

#### 1.5.4. Multipass and Slit Rheometry

Multipass rheometers and related slit rheometers [78] offer decoupling pressure and shear contributions to polymer relaxation. These rheometers utilize piston-driven reciprocating flow, facilitating the study of transient responses, stress birefringence, and relaxation dynamics under high pressure. Their capacity to operate with small sample volumes renders them well-suited for fundamental investigations of segmental mobility and time–pressure superposition. However, the technique is susceptible to artifacts such as compressibility, wall slip, and viscous heating. Appropriate data reduction protocols are crucial for obtaining reliable pressure coefficients. Similar to capillary rheometry, the shear rate, temperature, pressure, and shear stress are not uniform throughout the sample.

#### 1.5.5. Falling-Ball and Falling-Body Viscometers

For simpler fluids or near-Newtonian regimes, falling-ball viscometers [79,80] provide a mechanically straightforward method for determining viscosity under pressure. By monitoring the motion of a ball or cylinder through the fluid in a pressurized tube, the viscosity can be derived from the terminal velocity. Although they are appealing for their simplicity and compatibility with high pressures, these devices offer limited control over shear rates and are inadequately suited for non-Newtonian polymer melts.

#### 1.5.6. High-Pressure Sliding Plate Rheometry

A high-pressure sliding plate rheometer (HPSPR) is a unique scientific rheometer used to study rheological behavior of molten polymers at high-pressure. The HPSPR is a high-pressure version of a sliding plate rheometer (SPR) [67,68,81,82], which generates rectilinear deformation on a polymer melt sample by a use of an actuator while a shear stress transducer (SST) [83] measures the stress in the center of the sample. Thus, no torsional deformation is applied to the sample, unlike rotational rheometry, and there are no end or edge effects in the stress measurements. The rheometer is enclosed by a high-pressure vessel to construct the HPSPR [15,38,84] in which pressure increases by a hydraulic pump with an inert oil, and the HPSPR is placed in a convection oven. Hence, shear rate, temperature, pressure, stress are all uniform across the sample, controlled independently, and determined separately, in situ, with no corrections necessary since shear rate and pressure are controlled independently, and stress is measured independently of flow and pressure. The pressure used in the HPSPR is absolute hydrostatic pressure, and there is no gradient inside the rheometer. Its operating temperature and pressure ranges are from 25 to 225°C and a vacuum to 70 MPa. Its shear rate and stress range from $0.001$ to $500\ s^{-1}$ and from 0.1 to 600 kPa, respectively. Considering that polyethylenes show wall slip above 100 kPa, the HPSPR can also be used to study slip at high-pressure [44]. When supercritical fluid (SCF) is used as the pressurizing medium instead of inert oil, the effect of dissolved SCF on the rheological properties of polymers can also be studied [14]. In this study, we used the HPSPR to study the effects of pressure and molecular structures on the viscosity of polymer melts.

### 1.6. Effect of Branching Structure on Temperature and Pressure Sensitivity

The influence of molecular architecture, particularly short-chain branching (SCB) and long-chain branching (LCB), on the rheological behavior of polymers is a critical area of study in materials science [85,86] as it directly impacts processing conditions and end-use performance. Short-chain branching, which consists of branches with fewer than 10 carbons, is typically introduced through comonomers such as butene, hexene, or octene during polymerization with ethylene [58]. These branches significantly alter the polymer melt viscosity by influencing chain packing and compressibility [87]. Although SCB has not attracted as much attention as LCB [58,88,89], it has important effects on polymer rheology [58,90], making it crucial to understand how it affects the physical properties of polymers in response to pressure. While the branches in SCB are too short to entangle with neighboring chains, they instead disrupt regular chain alignment, which increases the initial free volume in the melt. This irregular structure makes the polymer more compressible, causing the free volume to change more rapidly with temperature variations. As a result, polymers with SCB exhibit steeper viscosity changes compared to their linear analogs of similar molecular weight. Additionally, the activation energy for flow increases with higher SCB content, branch density, and branch length at various temperatures [91]. Consequently, the way free volume expands or contracts with temperature changes is modified by SCB, leading to a different sensitivity, i.e. different activation energy for flow.

Long-chain branching has long captured the interest of polymer chemists and physicists [66,92] due to its significant effects on the flow behavior of polymers [93]. Long-chain branching refers to a structural feature where side chains attached to the main polymer backbone are long enough to participate in entanglements with the backbone or neighboring chains, typically exceeding the polymer's entanglement molecular weight from a physics perspective or having more than six carbons from an NMR perspective since differentiating the side-chain length becomes ambiguous after the fourth or fifth carbon. This type of branching is often introduced during polymerization processes, such as through the incorporation of multifunctional comonomers or via radical mechanisms at high pressure [94]. Consequently, polymers with LCB exhibit strain-hardening behavior [69,70,93,95], which is vital in polymer processes involving stretching, such as fiber spinning, blow molding, and foaming. And compared to linear or SCB polymers, those composed of LCB show much higher temperature sensitivities in their rheological properties [55,59,73,96,97] as LCB increases the activation energy for terminal relaxation, making the viscoelastic properties more responsive to thermal variations [98]. Additionally, LCB significantly increases the number of entanglements per molecule, leading to an increase in viscosity. These effects are attributed to heterogeneous relaxation modes as branched molecules require higher thermal energy for disentanglement. Moreover, LCB often induces thermorheologically complex behavior [90]. While the effect of branching on the temperature sensitivity of viscosity has been extensively studied, its impact on pressure sensitivity has received less attention [55]. In this study, we aim to explore the effect of branching on the pressure sensitivity of viscosity.

## 2. Experimental

### 2.1. Materials

Polyethylene (PE), the polymer with the simplest repeating structure, is widely used due to its low cost, chemical resistance, excellent processibility, toughness, and flexibility. For this work, four different polyethylenes with similar molecular weights, but distinct molecular architectures, were selected to investigate their rheological properties. The key properties of these polymers are summarized in Table 1. The first sample is a high-density polyethylene (HDPE, Japan Polyolefins) with no branching structure. HDPE is a commercial grade (J-REX HD KF251A) known to possess a broad molecular weight distribution. This HDPE grade is produced with a supported chromium or related transition-metal catalyst, which is known to yield very broad molecular-weight distributions (typically 8<PDI<20) [99]. Such broad distributions are desirable in industrial pipe and blow-molding grades because high-molecular-weight chains provide strength while lower-molecular-weight chains enhance processability. Linear low-density polyethylene (LLDPE, Dow Chemical Company) is a commercial ethylene-octene copolymer manufactured via a solution process and has a polydispersity of 3.8. The other two samples belong to the class of metallocene polyethylenes (mPEs), which are distinguished from traditional PEs like LLDPE, LDPE, and HDPE. Metallocene PEs are produced using single-site metallocene catalysts, in contrast to the multi-site Ziegler-Natta or metal oxide supported catalysts used for traditional linear polymers. This catalytic difference allows for a well-controlled molecular structure, such as a relatively low PDI. As a result, mPEs are valuable not only as commercial materials with superior toughness, optical properties, and heat-sealing characteristics, but also as ideal experimental materials. Two subclasses of mPE were used: a linear mPE (LmPE, Exxon Mobil) with short-chain branching (SCB) from a butene comonomer but no long-chain branching (LCB); and a long-chain branched mPE (BmPE, Dow Chemical Company) containing both LCB and SCB from an octene comonomer. The BmPE has a low level of long-chain branching, determined to be 0.06 LCB per 1,000 backbone carbons using the rheological and GPC-based technique developed by Wood-Adams and Dealy [96]. Carbon-13 NMR could not be used to determine the LCB level in the BmPE due to the inability of NMR spectroscopy to distinguish between the long branches and the six-carbon short branches that arise from the octene comonomer.

Table 1. Molecular and thermal properties determined at 0.1 MPa. $\eta_0$: zero-shear viscosity, $T_m$: melting temperature, $T_{m,h}$: melting temperature at highest peak of differential scanning calorimetry (DSC) thermogram, $T_{m,e}$: melting temperature at the end of peak of DSC thermogram, $M_w$: weight-average molecular weight, $M_n$: number average molecular weight, LCB: number of long-chain branching, and C: number of backbone carbon.

| Sample Code | Comonomer | $T_m$ (170°C) [100] | | Density (g/cc) at 20°C | $M_w$ (g/mol) | $M_w/M_n$ | LCB/$10^3$C | |
|---|---|---|---|---|---|---|---|---|
| | | $T_{m,h}$ | $T_{m,e}$ | | | | Rheology | GPC |
| HDPE [14] | | 130.5 | 134.4 | 0.945 | 111,000 | 13.6 | 0 | 0 |
| LLDPE [101] | Octene | 126 | 138 —140 | 0.923 | 119,600 | 3.8 | 0 | 0 |
| LmPE (L-mPE-3 [102], LDL1 [96]) | Butene 11.4 wt% (28.5 SCB/1000 C) | 105 | 112 —117 | 0.910 | 108,800 | 2.3 | 0 | 0 |
| BmPE (B-mPE-3 [102], (LDB3 [96]) | Octene | 104 | 113 | 0.908 | 89,400 | 2.3 | 0.053 [103] | 0.06 [103] |

## 2.2. Procedure

### 2.2.1. Sample Preparation

Prior to experiments, all polymer pellets samples were placed in a vacuum oven for over a week to remove any residual monomers or dissolved gases. Dried pellets were used for PVT measurements. For rheometry, samples were prepared by compression molding, a method chosen to minimize the strain and thermal history of the polymers. The pellets were placed into either rectangular (12 cm x 12 cm) or circular (25 mm diameter) molds with a 1 mm thickness. The molds were then set in a Carver press (USA), heated to 170°C, and a force of 3 tons was applied to shape the material into a uniform sheet. To minimize sample distortion, the mold was cooled with circulating water while the compressive force was maintained. The finished molded sheets were stored in a vacuum oven at room temperature until they were needed for testing. For the nonlinear rheological measurements, the rectangular sheets were cut into strips with dimensions of 2 cm x 10 cm x 1 mm.

### 2.2.2. Thermal Stability

To ensure that thermal degradation was negligible during the duration of the rheological measurements, the stability of each polymer was evaluated. The longest duration of high-pressure rheometry testing was five hours. The degree of degradation can be inferred by measuring the change in rheological properties over time. Time-sweep experiments using small-amplitude oscillatory shear (SAOS) were performed on an ARES rheometer (Rheometric Scientific). Each test was conducted for five hours at 170°C under a nitrogen atmosphere to prevent oxidative degradation. The experiments utilized parallel plates with a 25 mm diameter and a 1 mm gap, a strain of 5%, and a frequency of 1 rad/s. The magnitude of the complex viscosity for each sample increased by less than 5%. Based on this result, it was assumed that the polymer properties remained stable and that thermal degradation was insignificant during the rheological measurements.

### 2.2.3. Pressure–Volume–Temperature Behavior

The pressure–volume–temperature (PVT) properties of the samples were measured using a Gnomix PVT Apparatus [104]. The measurements were conducted over a temperature range of 30 to 200°C and a pressure range of 0 to 70 MPa, with data collected at 10 MPa intervals. Prior to each measurement, the sample chamber was evacuated and subsequently filled with mercury. The experiments were then performed isothermally by incrementally increasing the pressure. The resulting PVT data were described using Wohl's modified version of the empirical Tait equation (Eq. 1). This predictive model was essential in this study due to the rheological measurements being performed at pressures that differed from the specific pressures used in the PVT experiments. The model allowed for the accurate determination of the specific volume at the precise conditions of the rheological tests, which was then used to calculate the vertical shift factor (Eq. 5).

### 2.2.4. Non-Linear Rheometry at Various Pressures

A high-pressure sliding plate rheometer (HPSPR) was used to study the effect of pressure on the rheological properties of the three polyethylene samples. The interior of the rheometer was pressurized using a hand pump (Enerpac P-2282) with Krytox General Purpose Oil 107 (DuPont) as the pressurizing medium. This oil was chosen for its thermal stability, non-flammability, low volatility, and excellent lubrication properties, and because it does not swell polar or nonpolar materials. A sample strip was loaded onto one of the HPSPR plates, which had been pre-heated to 170°C, over the active face of the stress transducer. The rheometer was then sealed and pressurized to the desired setpoint. After a 30-minute thermal equilibration period, a series of steady simple shear tests were performed at various shear rates (0.03 to 70 s$^{-1}$) to obtain the stress growth function $[\sigma^+(t,\dot{\gamma},P)]$ as a function of shear rate and pressure by applying shear deformation until the steady state stress value $[\sigma(\dot{\gamma},P)]$ was obtained. A fresh sample was used to cover the full range of shear rates for each pressure condition. The highest shear rate was selected to ensure the stress did not exceed 100 kPa, which is the approximate onset stress for wall slip in these polymers [44]. The lowest shear rate was chosen so that the measured stress exceeded 2 kPa, the lower limit for reliable data from the transducer. Nonlinear rheological measurements were performed at four pressures: 0.1, 23, 46, and 69 MPa. To ensure data reliability, the experiment at the reference pressure ($P_0 = 0.1$ MPa) was replicated seven times while the measurements at the higher pressures were each replicated four times.

## 3. Results and Discussion

### 3.1. Effect of Pressure on Specific Volume and Vertical Shift Factor

The PVT measurements were performed from 30 to 200°C and 0 to 70 MPa in isothermal mode, and the data were fitted to the modified Tait model (Eq. 1), only for data in molten state. Table 2 shows the fitting parameters. Since the operating temperature of 170°C was of interest for this study, data only at this temperature shown in Figure 1. Figure 1(a) shows the specific volume data of each sample at 170°C up to 70 MPa, and the solid lines are modified Tait model predictions. Since there is no sudden drop of the specific volume with increasing pressure, all the samples were in molten state up to the maximum operating pressure. As can be predicted (Introduction section), the specific volume of the LCB polymer shows the highest value while HDPE without any branching shows the lowest value. LLDPE with SCB and a higher PDI shows higher specific volume than LmPE. Again, as can be predicted (Introduction section), SCB also increases the specific volume, and PDI increases specific volume as well.

To account for the change in the number of molecules per volume due to volume changes for rheological analyses, the vertical shift factor, $b_P(P)$, at various pressures (Eq. 5) is needed. The resulting values are plotted in Figure 1(b). The values at 23, 46, and 69 MPa are based on the modified Tait model predictions since no experimental PVT data were obtained at those pressures due to limited functions of the operating software for the Gnomix PVT instrument. Although the absolute value of the specific volume for BmPE is the highest [Figure 1(a)], surprisingly, $b_P(P)$ of all the samples (with or without SCB or LCB) shows a similar trend. Pressurizing to 70 MPa can increase the density by 6%. This much change is often ignored especially when time–pressure superposition is performed while such a change was covered by horizontal shift factor during time–pressure superposition, but the horizontal shift factor obtained this way will have inherent error by 6%. To avoid this issue, the vertical shift factors were used to obtain the reduced stress (Eq. 12) and then horizontal shift on the reduced stress was performed (Eq. 17) to obtained the horizontal shift factors (Eq. 19).

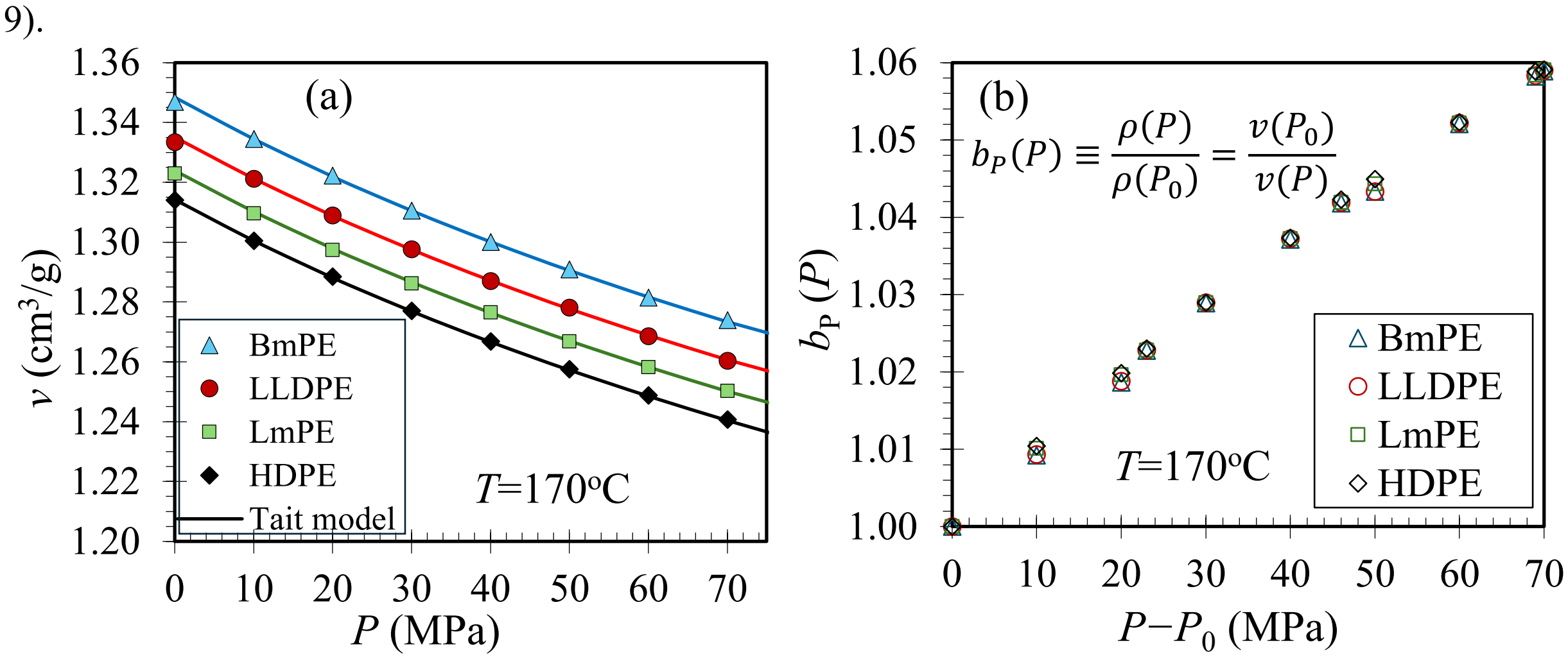

Figure 1. Pressure–volume–temperature (PVT) data for the four polyethylene samples at 170°C up to 70 MPa. (a) Specific volume as a function of pressure. Solid lines represent fits to the modified Tait equation. (b) Vertical shift factors ($b_P$)

derived from the PVT data. The values at 23, 46, and 69 MPa are predictions from the modified Tait model.

Table 2. Fitting parameters for the modified Tait equation used to describe the PVT data for each sample.

| | HDPE | LLDPE | LmPE | BmPE |
|---|---|---|---|---|
| $v_0$ (cc/g) | $8.76 \times 10^{-1}$ | $9.13 \times 10^{-1}$ | $9.08 \times 10^{-1}$ | $9.23 \times 10^{-1}$ |
| $v_1$ (cc/gK) | $9.84 \times 10^{-4}$ | $9.52 \times 10^{-4}$ | $9.39 \times 10^{-4}$ | $9.60 \times 10^{-4}$ |
| $B_0$ (MPa) | 677.03 | 549.60 | 636.35 | 550.07 |
| $B_1$ (1/K) | $4.80 \times 10^{-3}$ | $4.33 \times 10^{-3}$ | $4.66 \times 10^{-3}$ | $4.32 \times 10^{-3}$ |

### 3.2. Effect of Shear Rate on Viscosity

Shear-rate sweep tests were performed at 170°C and 0.1 MPa for the four polyethylene samples. To ensure data reliability, experiments were repeated seven times for each material, with the maximum relative standard deviation not exceeding 5%. Figure 2 presents viscosity ($\eta$) versus shear rate ($\dot{\gamma}$). Solid lines represent Cross model fitting (Eq. 8), and error bars are omitted from the plots to avoid confusion. Table 3 summarizes the corresponding parameters. All samples exhibit the characteristic shear-thinning behavior of polymer melts, but the degree of shear thinning differs markedly across samples due to variations in molecular structure. These results highlight how different molecular features (PDI, SCB, and LCB) contribute differently to shear thinning.

Long-chain branching (LCB) strongly enhances shear thinning, as seen in BmPE (octene comonomer, LCB, and low PDI). The absence of a Newtonian plateau is observed within the experimental window, consistent with the review in Introduction section. The broad molecular weight distribution (PDI = 13.6) of HDPE also induces strong shear thinning, comparable in magnitude to the effect of LCB in BmPE, dominated by broad relaxation spectrum [105]. Although HDPE has $M_{\mathrm{w}}$ similar to that of the other polymers, such a high PDI implies that there are both more shorter and more longer chains than the other polymers, with the longer chains especially contributing to the high zero-shear viscosity ($\eta_0$). For both materials, determining the zero-shear viscosity is challenging because the Newtonian regime lies outside the accessible shear-rate range. However, an experimental protocol for extrapolation has been described by Park and Dealy [14]. By contrast, linear metallocene polyethylene with a narrow distribution (LmPE, butene comonomer, low PDI) exhibits the weakest shear thinning and a well-defined Newtonian plateau while the structure with SCB appears to moderate the influence of PDI on shear thinning with LLDPE (octene comonomer and moderate PDI) falling in between, showing modest shear thinning. However, the presence of SCB introduces local packing disruptions that compact the coil and reduce chain extensibility under flow, thereby diminishing the sensitivity of shear thinning to PDI.

At very high shear rates, the viscosity of polymer melts often converges toward similar values if their microstructures are identical since entanglement effects are largely suppressed and flow resistance is governed by intrachain friction and backbone dynamics [85]. However, in the present study the viscosity curves of HDPE, LLDPE, LmPE, and BmPE remain distinct even at the highest shear rates investigated. This divergence indicates that their microstructural differences, specifically, the type and distribution of comonomers (octene vs. butene) and presence or absence of SCB, continue to influence chain orientation, residual intermolecular interactions, and flow resistance under strong deformation. Thus, the lack of merging at high $\dot{\gamma}$ suggests that the flow behavior at high rate is not universal across these samples but instead reflects persistent microstructural effects. In particular, comonomer identity controls local stiffness and packing; and broad polydispersity introduces faster relaxing short-chains that can lubricate flow differently. These effects prevent collapse of the viscosity curves to a common asymptote, reinforcing that rheological measurements can serve as a sensitive probe of microstructural distinctions among nominally similar polyethylenes.

The viscosity crossover observed around the characteristic shear rate $\dot{\gamma}_C = 4\ \mathrm{s}^{-1}$ ($\eta = 6.6$ kPa s and $\sigma = 26.4$ kPa) for HDPE, LLDPE, and LmPE, with BmPE approaching the same point. This crossover suggests that at $\dot{\gamma}_C$ the balance of competing structural effects of molecular weight, molecular-weight polydispersity, SCB, and LCB converges to yield similar flow resistance. In terms of time scales, $\dot{\gamma}_C$ can be interpreted as the inverse of a characteristic relaxation time, $\tau^* \approx 1/\dot{\gamma}_C$, associated with the common backbone molecular-weight scale of these samples. Because all four grades have similar $M_w$ (Table 2), their longest relaxation times and the onset of strong shear thinning are expected to occur at comparable rates even though the detailed shapes of their flow curves differ due to branching and PDI. Below $\dot{\gamma}_C$, BmPE shows the highest viscosity due to the motion restriction due to the branching points [59] while LLDPE and LmPE show lower values depending on PDI and comonomer type. At the crossover, strong shear thinning in BmPE and HDPE collapses their viscosities to values comparable with those of LLDPE and LmPE. Above $\dot{\gamma}_C$, viscosities diverge again, reflecting differences in smaller scale structure such as chain alignment, residual friction, and comonomer effects. At sufficiently high shear rates, BmPE even falls below the linear polymers due to enhanced shear-thinning efficiency.

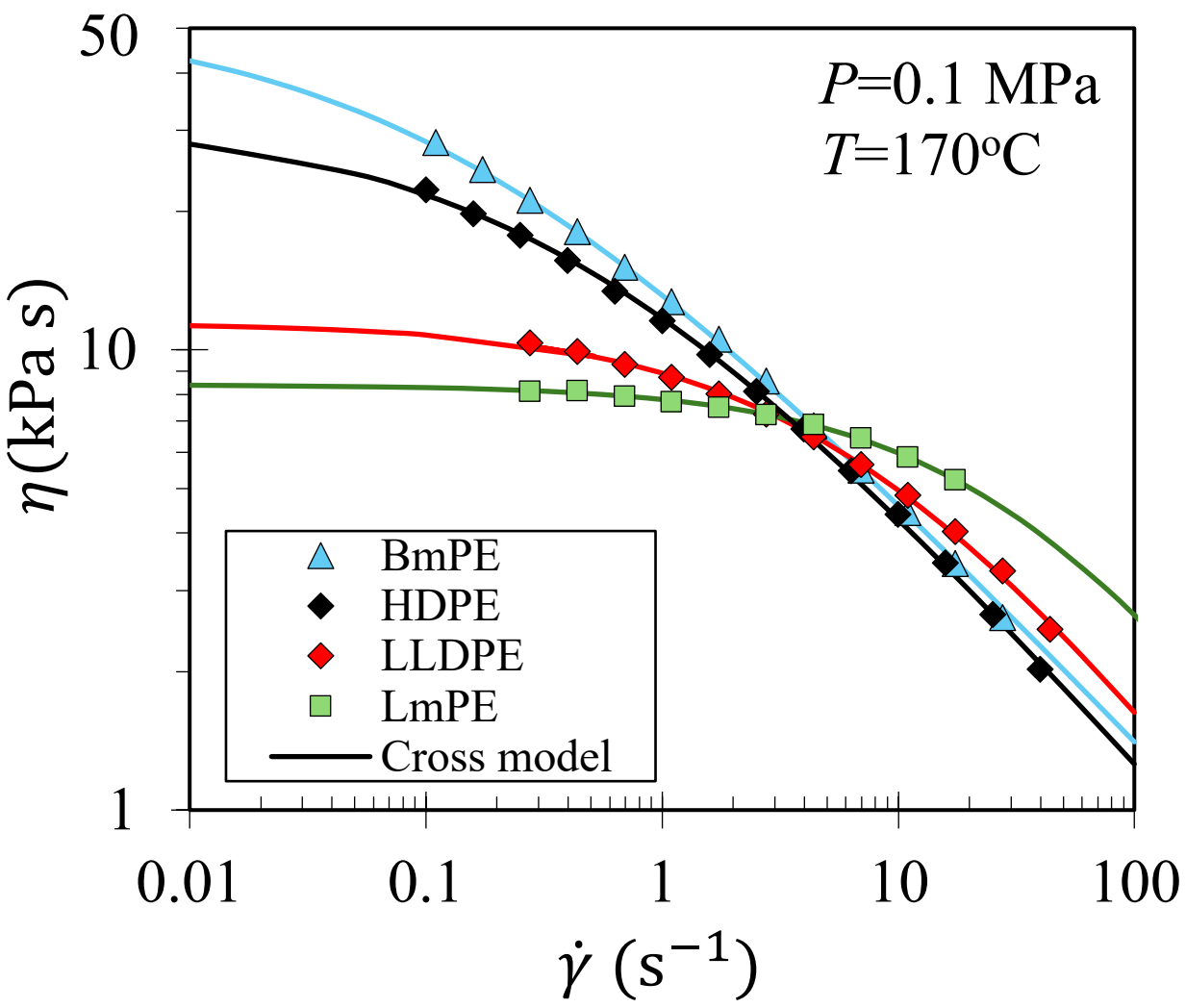

Figure 2. Viscosity as a function of shear rate for the four polyethylene samples at the reference condition of 170°C and 0.1 MPa. Solid lines represent fits to the Cross model. The data shown are the average of seven replicate measurements. The relative standard deviation for all points was 5% or lower. For visual clarity, error bars are not displayed

Table 3. Cross model fitting parameters, horizontal pressure shift factors, and the Barus model pressure–viscosity coefficient.

| $P$ (MPa) | HDPE | | | LLDPE | | | LmPE | | | BmPE | | |
|---|---|---|---|---|---|---|---|---|---|---|---|---|
| | $\beta = 10.87\ \mathrm{GPa}^{-1}$ | | | $\beta = 18.67\ \mathrm{GPa}^{-1}$ | | | $\beta = 18.44\ \mathrm{GPa}^{-1}$ | | | $\beta = 25.00\ \mathrm{GPa}^{-1}$ | | |
| | $m = 0.577$ | | | $m = 0.662$ | | | $m = 0.725$ | | | $m = 0.541$ | | |
| | $\eta_0$(kPa s) | $\lambda$(s) | $a_P$ | $\eta_0$(kPa s) | $\lambda$(s) | $a_P$ | $\eta_0$(kPa s) | $\lambda$(s) | $a_P$ | $\eta_0$(kPa s) | $\lambda$(s) | $a_P$ |
| 0.1 | 31.29 | 2.43 | 1.000 | 11.4 | 0.15 | 1.00 | 8.40 | 0.029 | 1.000 | 53.2 | 7.85 | 1.000 |
| 23 | 41.05 | 3.19 | 1.312 | 18.2 | 0.22 | 1.58 | 13.48 | 0.045 | 1.559 | 92.2 | 14.50 | 1.846 |
| 46 | 50.48 | 3.92 | 1.613 | 23.3 | 0.30 | 2.47 | 20.34 | 0.067 | 2.230 | 159.42 | 23.53 | 3.000 |
| 69 | 66.87 | 5.20 | 2.137 | 42.9 | 0.56 | 3.74 | 32.13 | 0.104 | 3.575 | 281.27 | 41.52 | 5.287 |

### 3.3. Effect of Pressure on Viscosity

To determine the effect of pressure on viscosity, steady shear experiments were conducted for each polymer at 170°C and at 23, 46, and 69 MPa. To ensure data reliability, experiments were repeated four times for each pressure and material, with the maximum relative standard deviation not exceeding 5%. Figure 3 shows the effect of pressure on $\sigma(P)$ or $\eta(P)$ versus $\dot{\gamma}$ (open symbols) for HDPE. Error bars are omitted from the plots to avoid confusion. Pressure shifts the stress curves primarily to the left (to lower shear rates) and slightly upward. This diagonal shift occurs because pressure slows the time response of the polymer (a horizontal shift) and also increases its density, which slightly increases the stress (a vertical shift). The viscosity curves at high pressure are shifted upward and to the left, which is in the diagonal direction since the viscosity has the shear rate in its denominator, which is affected by pressure.

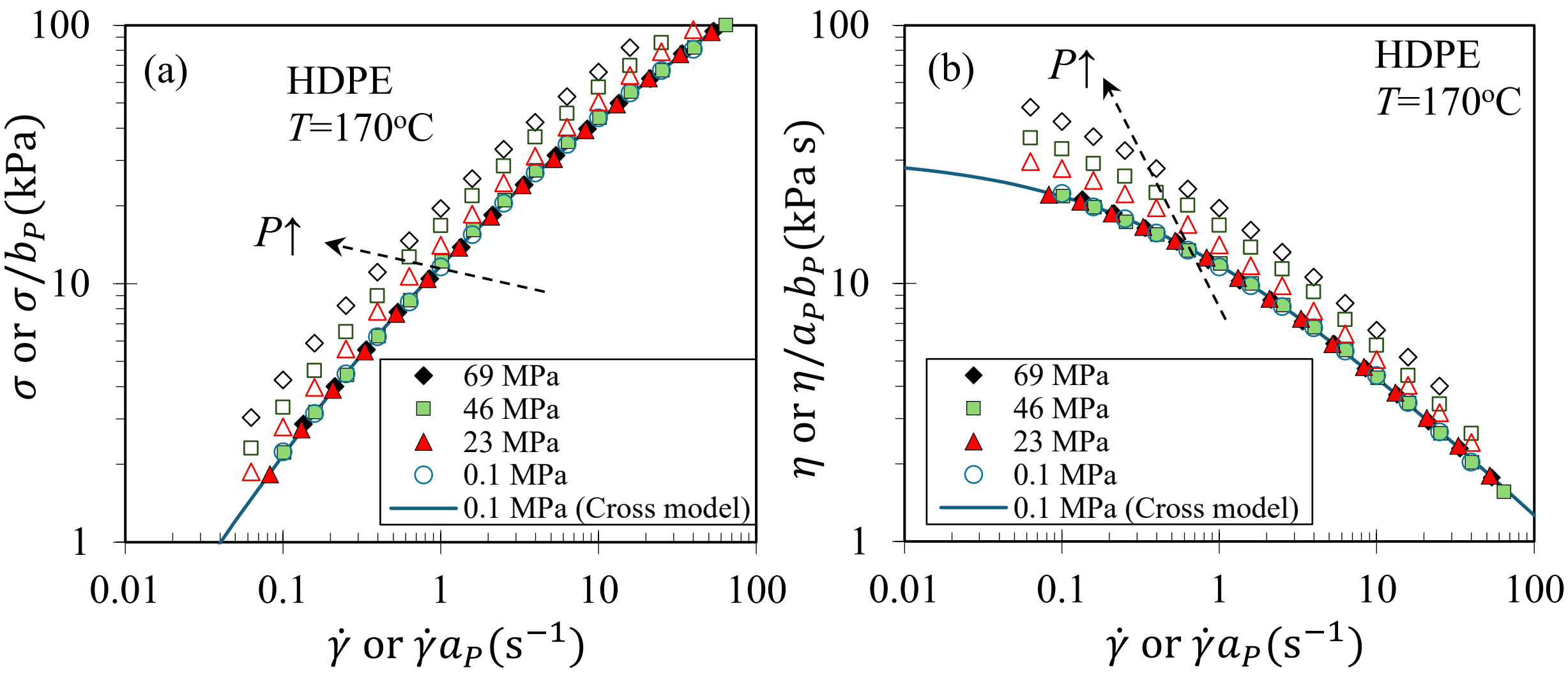

Figure 3. Rheological behavior of HDPE at 170°C and at various pressures. The data shown are the average of four replicate measurements. The relative standard deviation for all points was 5% or lower. For visual clarity, error bars are not displayed. (a) Shear stress and (b) viscosity as a function of shear rate. Open symbols represent the raw data at each pressure. Solid symbols represent the data shifted to the reference pressure of 0.1 MPa to construct the master curve: $[\sigma(P)/b_P(P)$ versus $\dot{\gamma}a_P(P)$ or $\eta(P)/a_P(P)b_P(P)$ versus $\dot{\gamma}a_P(P)]$. The solid line is the Cross model fit to the reference data. Symbols for 0.1 MPa data and shifted 46 MPa data are almost overlapped so those are not clearly differentiated on the plots.

To quantify effects of pressure, time–pressure superposition was applied. First, the horizontal shift factor $a_P(P)$ was determined by shifting the reduced stress, $\frac{\sigma(\dot{\gamma},P)}{b_P(P)}$ (Eq. 12) data onto a single master curve at $P_0 = 0.1$ MPa. Figure 3(a) also shows both horizontally and vertically shifted stress $[\sigma(P)/b_P(P)$ versus $\dot{\gamma}a_P(P)$, Eq. 17] (solid symbols) for HDPE. Isolating the reduced stress allows for a direct determination of the horizontal shift as it requires shifting in only one dimension. The optimization of this shift was guided by minimizing the deviation from the Cross model prediction at 0.1 MPa since there were no overlapping data points at the reference pressure of 0.1 MPa. The resulting horizontal pressure shift factors, $a_P(P)$ for all samples are listed in Table 3. Once $a_P(P)$ was determined by shifting the reduced stress, the final viscosity master curve (Eq. 16**Error! Reference source not found.**) was constructed. Figure 3(b) shows r esulting master curves of the viscosity $[\eta(P)/a_P(P)b_P(P)$ versus $\dot{\gamma}a_P(P)]$ (solid symbols) for HDPE. This same procedure was applied to the other polymers, with their respective raw data and master curves presented in Figure 4 through Figure 6.

Two key features are common to all four samples. Firstly, the rheological data for each polymer was able to be successfully superposed onto a single master curve using a single pressure shift factor, $a_P(P)$ for each pressure. This indicates that all four samples, regardless of their molecular structure, exhibit piezorheologically simple behavior [64,65]. This finding implies that neither short-chain branching (SCB) nor long-chain branching (LCB) induces piezorheological complexity in polyethylene, a behavior that has also been observed in long-chain branched polystyrene (LCB-PS) [55]. This piezorheological simplicity contrasts with the known thermorheological complexity of the BmPE sample. Wood-Adams and Costeux [73] reported that this same polymer (identified as LDB3 in their study) fails time–temperature superposition because its relaxation modes have different temperature dependencies. This raises a critical question: why can a polymer be thermorheologically complex but piezorheologically simple? The answer may lie in the different physical mechanisms by which temperature and pressure affect polymer dynamics. Temperature affects molecular motion by providing thermal energy to overcome local energy barriers (activation energy, $E_a$). In a long-chain branched polymer like BmPE, distinct relaxation mechanisms exist, such as the different relaxation of the backbone and the branches. These modes have different activation energies and because the temperature dependence of a relaxation time is exponential with activation energy, a change in temperature shifts these modes by different amounts. This differential shifting breaks the superposition principle, leading to thermorheological complexity. Meanwhile, pressure, in contrast, primarily affects molecular motion by reducing the available free volume and increasing friction among chains. This acts as a global "brake", regardless of whether they are part of a branch or the backbone. The pressure dependence of relaxation is related to an activation volume, which corresponds to the size of the rearranging segment. Since this segmental size is roughly the same throughout the polymer, pressure scales all relaxation times by the same factor, $a_P(P)$. This uniform shifting preserves the shape of the relaxation spectrum, resulting in piezorheological simplicity.

A second key finding is that pressure does not alter the shear-thinning behavior of the polymers. (Indeed, this phenomenon is directly related to piezorheologically simple behavior). The successful superposition using only one horizontal shift factor, $a_P(P)$, at each pressure implies that the parameters for the Cross model (Eq. 13), $\eta_0(P_0)$, $\lambda(P_0)$, and $m(P_0)$ determined at 0.1 MPa ($P_0$), which define the shape of the flow curve, are independent of pressure. While shear-thinning behavior is strongly dictated by molecular structure (SCB, LCB, and PDI), pressure appears to uniformly scale the relaxation times of materials without modifying their relative distribution. In practical terms, the mechanisms of shear thinning, such as chain alignment and disentanglement, follow the same pathway regardless of the applied pressure; pressure only changes the overall timescale on which these events occur.

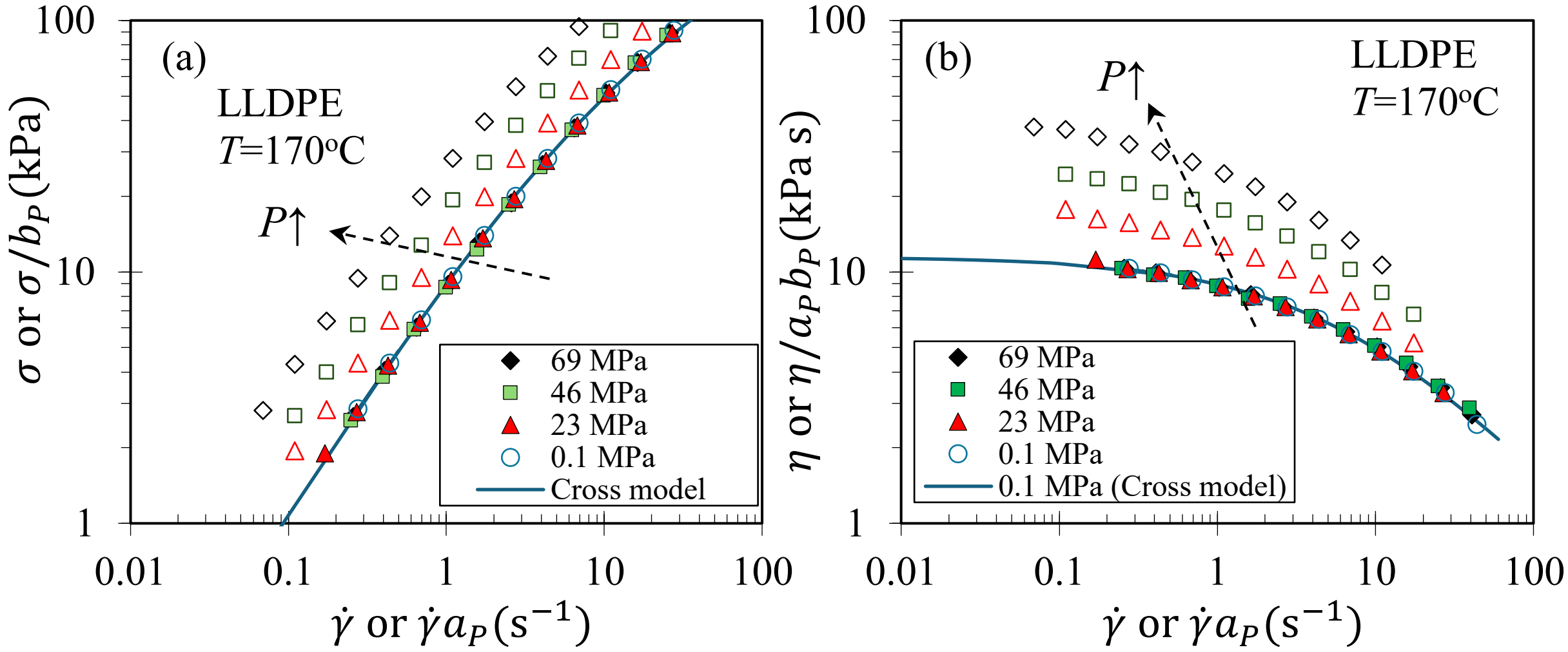

Figure 4. Rheological behavior of LLDPE at 170°C and at various pressures. The data shown are the average of four replicate measurements. The relative standard deviation for all points was 5% or lower. For visual clarity, error bars are not displayed. (a) Shear stress and (b) viscosity as a function of shear rate. Open symbols represent the raw data at each pressure. Solid symbols represent the data shifted to the reference pressure of 0.1 MPa to construct the master curve: $\sigma(P)/b_P(P)$ versus $\dot{\gamma}a_P(P)$ or $\eta(P)/a_P(P)b_P(P)$ versus $\dot{\gamma}a_P(P)$].

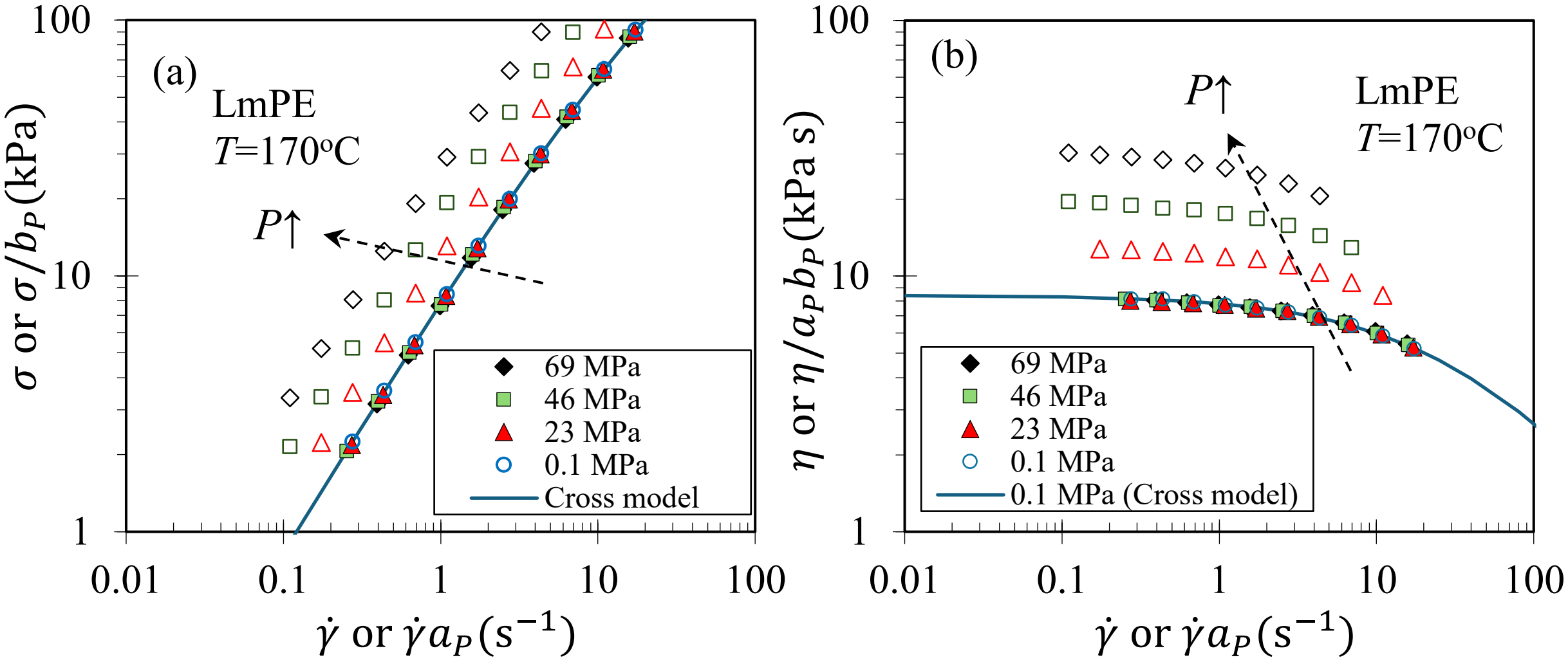

Figure 5. Rheological behavior of LmPE at 170°C and at various pressures. The data shown are the average of four replicate measurements. The relative standard deviation for all points was 5% or lower. For visual clarity, error bars are not displayed. (a) Shear stress and (b) viscosity as a function of shear rate. Open symbols represent the raw data at each pressure. Solid symbols represent the data shifted to the reference pressure of 0.1 MPa to construct the master curve:

$[\sigma(P)/b_P(P)$ versus $\dot{\gamma}a_P(P)$ or $\eta(P)/a_P(P)b_P(P)$ versus $\dot{\gamma}a_P(P)]$.

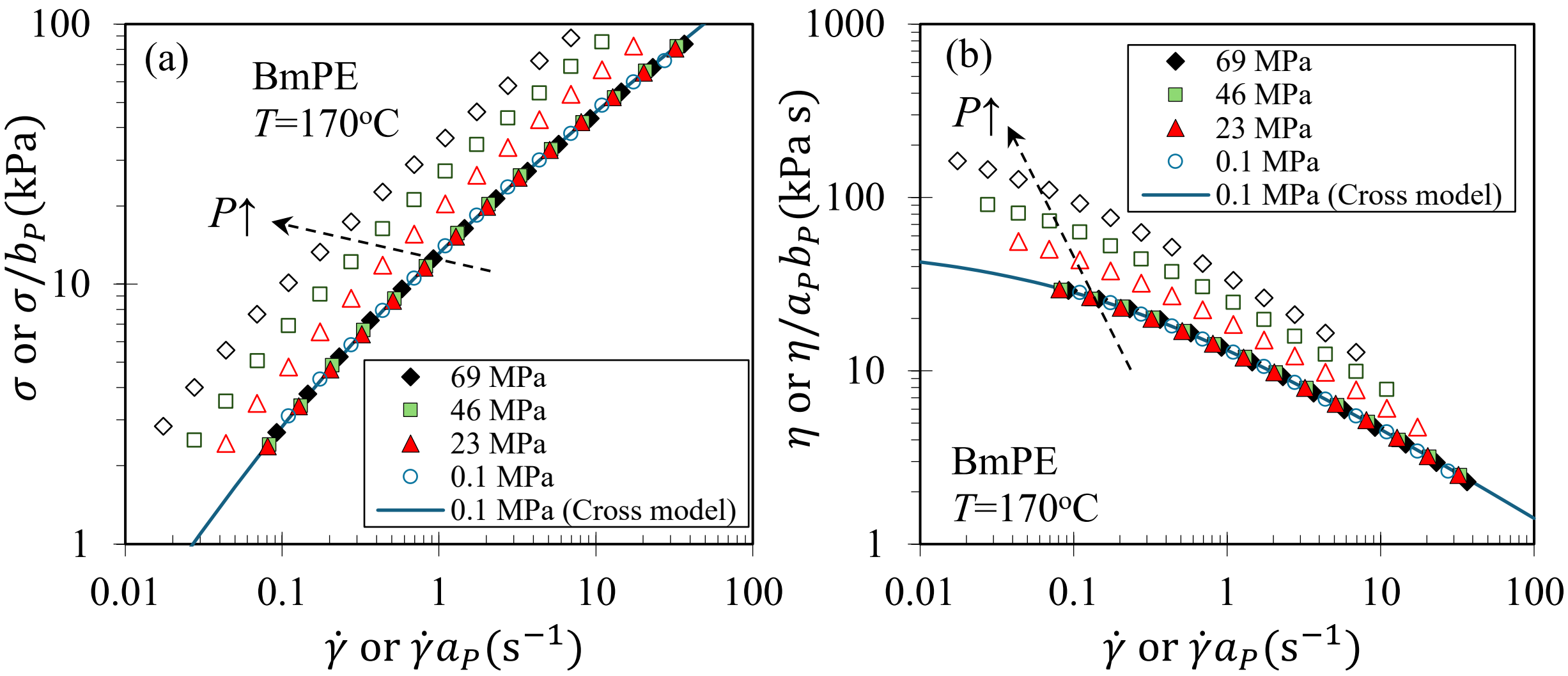

Figure 6. Rheological behavior of BmPE at 170°C and at various pressures. The data shown are the average of four replicate measurements. The relative standard deviation for all points was 5% or lower. For visual clarity, error bars are not displayed. (a) Shear stress and (b) viscosity as a function of shear rate. Open symbols represent the raw data at each pressure. Solid symbols represent the data shifted to the reference pressure of 0.1 MPa to construct the master curve: $[\sigma(P)/b_P(P)$ versus $\dot{\gamma}a_P(P)$ or $\eta(P)/a_P(P)b_P(P)$ versus $\dot{\gamma}a_P(P)]$. The solid line is the Cross model fit to the reference data.

### 3.4. Effect of Molecular Structure on Pressure Sensitivity

A natural follow-up question is how the pressure sensitivity of the viscosity can be quantified and how this sensitivity is affected by molecular structure. To address this, the Barus model (Eq. 20) was used to determine the pressure–viscosity coefficient ($\beta$) from the pressure shift factors ($a_P$). This coefficient serves as a quantitative measure of a pressure sensitivity of viscosity. Figure 7 shows the shift factors as a function of pressure, where the slope of the line is the pressure coefficient, $\beta$, of which values are listed in Table 3. The results clearly show that the BmPE, which contains both LCB and SCB, has the highest pressure sensitivity ($\beta = 25.1$ GPa$^{-1}$). In contrast, the two linear polymers with different polydispersity, LLDPE and LmPE, have very similar and lower pressure coefficients ($\beta \approx 19.5$ GPa$^{-1}$). Thus, we can see that any (either short or long) chain branching structure increases pressure sensitivity. The side groups act as obstacles to intermolecular motion, and under compression, this steric hindrance becomes more pronounced, leading to a sharper increase in viscosity given pressure change. However, LCB increases the pressure sensitivity much more than SCB. Polydispersity has a negligible effect on the pressure coefficient in these polymers. Despite the significant difference in PDI between LLDPE (3.8) and LmPE (2.1), their pressure sensitivities are nearly identical. Although HDPE has much higher PDI than these two linear polymers, HDPE shows the lowest pressure sensitivity while the high PDI increased the shear-thinning behavior significantly (Figure 2). The successful application of the Barus model, which is based on the absolute reaction rate theory, suggests that the viscosity increase under pressure is primarily due to an increase in the energetic cost of molecular motion, i.e. the response to time was slowed down. This contrasts with the free volume theory, which posits that viscosity increase is mainly because the available space for a polymer segment to move into disappears under compression. Our results imply that for these polymers at 170°C, there is sufficient free volume, and the main barrier to flow is the energy required for a segment to overcome the heightened intermolecular resistance in the compressed state.

To place these findings in a broader context, it is useful to consider polymers with more bulky and regular side groups, such as polystyrene (PS), which can be viewed as polyethylene with a phenyl group on every other carbon. Park et al. [55] studied the rheology of linear and long-chain branched (four-arm star) polystyrene and reported several striking results. Firstly, they found that the temperature sensitivity of PS viscosity was best described by the WLF equation (free volume theory) while the pressure sensitivity was best described by the Barus equation (reaction rate theory). This confirms that temperature and pressure affect viscosity through fundamentally different mechanisms. Secondly, the pressure coefficient for linear PS was $\beta = 31$ GPa$^{-1}$, which is significantly higher than the $\beta \approx 25$ GPa$^{-1}$ that we observed for BmPE. This demonstrates that the large, regular phenyl side groups in polystyrene have a much stronger impact on pressure sensitivity than the less frequent SCB and LCB in our polyethylene samples. Thirdly, the $\beta$ value for their LCB polystyrene was nearly identical to that of their linear PS. This suggests that once a polymer possesses a high degree of steric hindrance from large, regular side groups, the additional effect of long-chain branching on pressure sensitivity becomes minor. In other words, the phenyl groups significantly increase the pressure sensitivity while the LCB does not add significantly to this effect.

This study highlights the dominant role of side branches on the pressure sensitivity of viscosity. Future work could expand on these findings by investigating polymers with other types of structural variations. For example, comparing linear and LCB polypropylene would reveal the comparison between the effect of regular, short methyl side groups and that of LCB. Another interesting avenue would be to study polymers with stiffening groups with phenylene within the backbone, such as polycarbonate (PC), to understand how chain rigidity affects the pressure sensitivity of linear and branched PC.

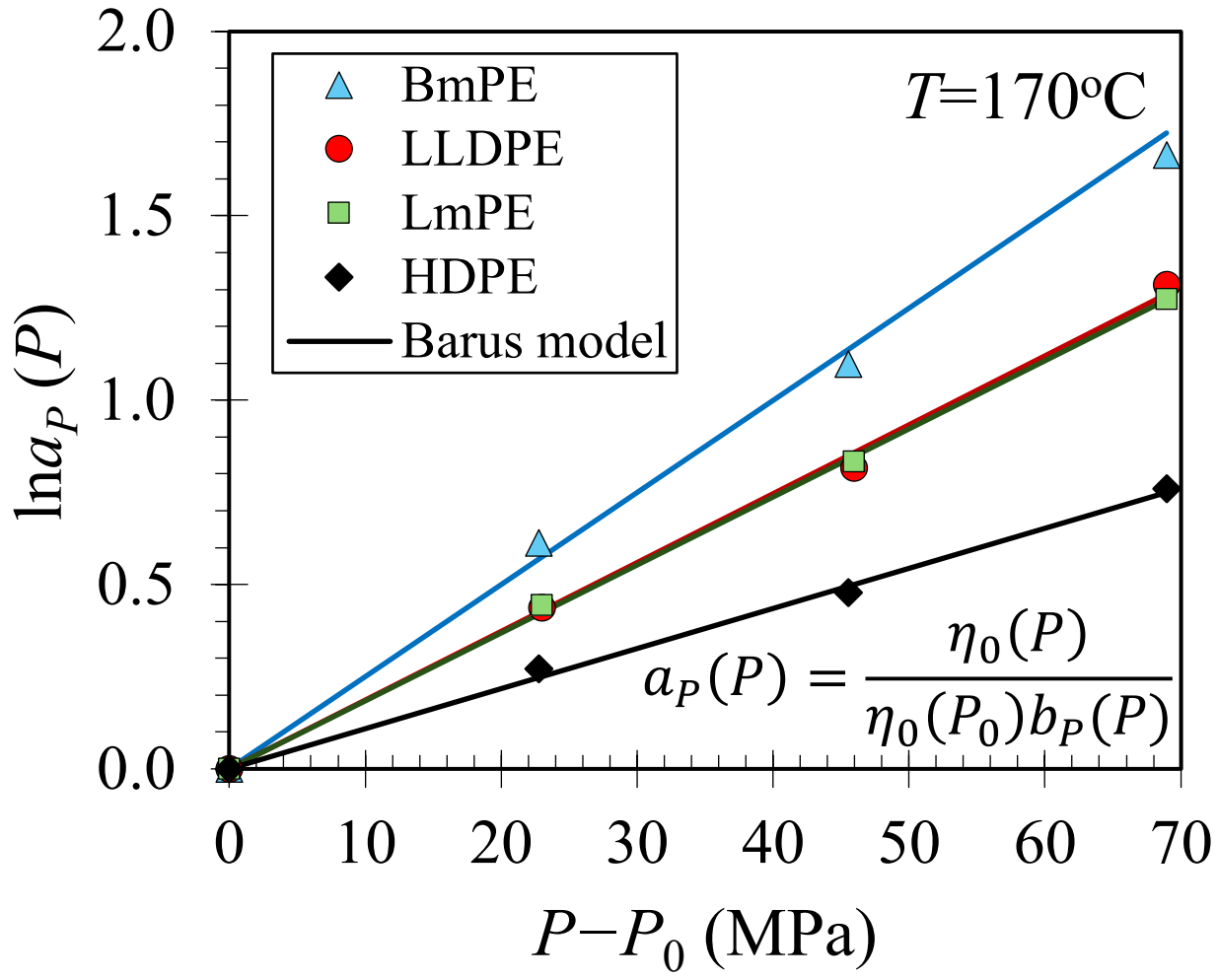

Figure 7. Horizontal pressure shift factor ($a_P$) as a function of pressure for each polyethylene sample. The solid lines represent linear fits according to the Barus model Eq. 20, from which the pressure–viscosity coefficient ($\beta$) is determined as the slope.

## 4. Conclusions

This study investigated the effects of pressure and molecular structure, specifically short-chain branching (SCB), long-chain branching (LCB), and polydispersity (PDI), on the rheological behavior of four distinct polyethylene melts employing a high-pressure sliding plate rheometer. The extent of shear thinning was directly linked to molecular architecture. The presence of LCB (in BmPE) and high PDI (in HDPE) both significantly enhanced shear-thinning behavior, making it difficult to determine their zero-shear viscosity. In contrast, the linear polymer with a low PDI (LmPE) exhibited the weakest shear thinning and a clear Newtonian plateau. Despite their structural differences, all four polymers exhibited piezorheologically simple behavior. Viscosity data at various pressures could be superposed onto a single master curve using one horizontal shift factor over the entire shear rate ranges. This indicates that pressure uniformly scales all relaxation times without altering the fundamental shape of the flow curve or the nature of shear thinning. The BmPE sample, known to be thermorheologically complex, followed time–pressure superposition perfectly, i.e., a polymer can be piezorheologically simple yet thermorheologically complex. This distinction arises because temperature differentially affects relaxation modes with different activation energies while pressure acts as a global brake shifting all relaxation modes by the same factor. The pressure sensitivity of viscosity (showing how much the viscosity increases given pressure change), quantified by the pressure-viscosity coefficient ($\beta$), is strongly dependent on branching but not on polydispersity. Both SCB and LCB increase the pressure sensitivity, with the BmPE sample showing the highest $\beta$ value. Side branches act as steric hindrances that become more pronounced under compression, leading to a sharper increase in viscosity. However, such an increase in the pressure coefficient for BmPE is much smaller than that for linear and branched PS, implying that the regular, bulky side groups have much stronger effects on the pressure sensitivity than LCB. On the other hand, PDI had little effect on the pressure coefficient in the polymers studied. In summary, while molecular structure dictates the fundamental shear-thinning behavior of a polymer melt, pressure acts as a uniform scaling factor on its dynamics. The sensitivity to this scaling is primarily enhanced by the presence of molecular branches, a finding critical for the accurate modeling and simulation of high-pressure polymer processing.

## Acknowledgments

The authors gratefully acknowledge the Department of Chemical and Paper Engineering at Western Michigan University (USA) for its partial financial support. We extend our sincere thanks to Prof. Musa Kamal and Dr. Nitin Borse at McGill University (Canada) for providing access to and training on the Gnomix PVT instrument. We are also indebted to Prof. Paula Wood-Adams at the University of Northern British Columbia (Canada), Dr. Seungoh Kim at McGill University (Canada), and Mr. Yoshito Sasaki at Japan Polyolefins (Japan) for generously sharing samples and their priceless advice. H.E.P. would also like to express appreciation to Prof. A. Jeffrey Giacomin at Queen's University (Canada), Dr. Chunxia He at Dow Chemical (USA), Dr. Stéphane Costeux at DuPont (USA), Prof. Benoît Blottière at Université Jean Monnet Saint-Etienne (France), Dr. Stewart McGlashan at Industrial Energy Systems (Australia), Dr. Yong-Gyun Cho at Cygteco (Korea), and Prof. Byoung-Uk Cho at Kangwon National University (Korea) for their invaluable guidance. Most importantly, H.E.P. offers his deepest and most profound gratitude to his mentor, Prof. John Michael Dealy of McGill University (Canada). Prof. Dealy's contributions to this work and to H.E.P.'s development as a scientist have been immeasurable. His unwavering guidance, financial support, and inspirational mentorship provided the foundation for this research. Prof. Dealy meticulously shaped H.E.P.'s scientific methodology, from fundamental problem-solving strategies to the specific writing style reflected in this manuscript. Beyond the realms of polymer physics and rheology, he imparted a deeper understanding of the scientific process and essential transferable skills, including effective presentation, professional communication, and cross-cultural awareness, and all of these contribute to this study. The opportunity to work within his leading research group, alongside world-renowned scientists, was a formative experience. Prof. Dealy's relentless efforts to help H.E.P. build a professional network have created a priceless asset for his future career, and again these led to the outcomes of this study. His influence is a debt that can never be fully repaid.

## Data availability

The data that supports the findings of this study are available within the article.

## References

1. Lin, L.; Lee, Y.; Park, H.E. Recycling and rheology of poly(lactic acid) (PLA) to make foams using supercritical fluid. *Physics of Fluids* **2021**, *33*, doi:10.1063/5.0050649.
2. An, H.-J.; Park, H.E.; Cho, B.-U. Effect of Temperature of Tetraethylammonium Hydroxide/Urea/Cellulose Solution on Surface Tension and Cellulose Bead Size. *Journal of Korea Technical Association of The Pulp and Paper Industry* **2021**, *53*, 69-76, doi:10.7584/jktappi.2021.12.53.6.69.
3. Xu, F.; Park, H.E.; Cho, B.-U. Porous carboxymethylcellulose/polyethyleneimine composite beads: formation process, enhanced physical properties, and pH-induced response mechanism. *Cellulose* **2024**, *31*, 6295-6316, doi:10.1007/s10570-024-05987-6.
4. Egerton, S.; Sim, C.; Park, H.E.; Staiger, M.P.; Patil, K.M.; Cowan, M.G. Controlling coefficients of thermal expansion in thermoplastic materials: effects of zinc cyanide and ionic liquid. *Materials Advances* **2022**, *3*, 4155-4158, doi:10.1039/d2ma00338d.
5. Lin, L.; Dang, Q.A.; Park, H.E. Enhanced Degradability, Mechanical Properties, and Flame Retardation of Poly(Lactic Acid) Composite with New Zealand Jade (Pounamu) Particles. *Polymers* **2023**, *15*, doi:10.3390/polym15153270.
6. Lin, L.; Mirkin, S.; Park, H.E. Biodegradable Composite Film of Brewers' Spent Grain and Poly(Vinyl Alcohol). *Processes* **2023**, *11*, doi:10.3390/pr11082400.
7. Lin, L.; Joe, M.; Dang, Q.A.; Park, H.E. Enhancing Biodegradation of Poly(lactic acid) in Compost at Room Temperature by Compounding Jade Particles. *Polymers* **2025**, *17*, doi:10.3390/polym17152037.
8. Zuliki, M.; Zhang, S.; Tomkovic, T.; Hatzikiriakos, S.G. Capillary flow of sodium and zinc ionomers. *Physics of Fluids* **2020**, *32*, doi:10.1063/1.5145303.
9. Tyagi, P.K.; Kumar, R.; Mondal, P.K. A review of the state-of-the-art nanofluid spray and jet impingement cooling. *Physics of Fluids* **2020**, *32*, doi:10.1063/5.0033503.
10. Ma, J.; Huang, Q.; Zhu, Y.; Xu, Y.-Q.; Tian, F.-B. Effects of fluid rheology on dynamics of a capsule through a microchannel constriction. *Physics of Fluids* **2023**, *35*, doi:10.1063/5.0165614.

11. Naseri, H.; Koukouvinis, P.; Malgarinos, I.; Gavaises, M. On viscoelastic cavitating flows: A numerical study. *Physics of Fluids* **2018**, *30*, doi:10.1063/1.5011978.
12. McIlroy, C. A fundamental rule: Determining the importance of flow prior to polymer crystallization. *Physics of Fluids* **2019**, *31*, doi:10.1063/1.5129119.
13. Kanso, M.A.; Giacomin, A.J. Van Gurp–Palmen relations for long-chain branching from general rigid bead-rod theory. *Physics of Fluids* **2020**, *32*, doi:10.1063/5.0004513.
14. Park, H.E.D., J.M. Effects of pressure and supercritical fluids on the viscosity of polyethylene. *Macromolecules* **2006**, *39*, 5438-5452, doi:10.1021/ma060735+.
15. Rudolph, N.M.; Agudelo, A.C.; Granada, J.C.; Park, H.E.; Osswald, T.A. WLF model for the pressure dependence of zero shear viscosity of polycarbonate. *Rheologica Acta* **2016**, *55*, 673-681, doi:10.1007/s00397-016-0945-4.
16. Bair, S.S. *High Pressure Rheology for Quantitative Elastohydrodynamics*, 2nd ed.; Elsevier: 2019.
17. Levitas, V.I. *Large Deformation of Materials With Complex Rheological Properties at Normal and High Pressure*; Nova Science Pub Inc: 1996.
18. Liu, J.; Zhao, B.; Kontziampasis, D.; Jiang, B.; Wu, W. A novel method for determining the pressure dependent characteristics of polymer melt during micro injection molding. *Polymer Testing* **2024**, *137*, doi:10.1016/j.polymertesting.2024.108520.
19. Housiadas, K.D.; Gryparis, E.; Georgiou, G.C. Annular Newtonian Poiseuille flow with pressure-dependent wall slip. *European Journal of Mechanics - B/Fluids* **2025**, *109*, 299-308, doi:10.1016/j.euromechflu.2024.10.012.
20. Burger, N.A.; Loppinet, B.; Clarke, A.; Petekidis, G. Pressure-Dependent Yield Stress of Organoclay-Based Gels. *Langmuir* **2025**, *41*, 20638-20647, doi:10.1021/acs.langmuir.5c02052.
21. Park, H.E.; Gasek, N.; Hwang, J.; Weiss, D.J.; Lee, P.C. Effect of temperature on gelation and cross-linking of gelatin methacryloyl for biomedical applications. *Physics of Fluids* **2020**, *32*, doi:10.1063/1.5144896.
22. Pouliot, R.A.; Young, B.M.; Link, P.A.; Park, H.E.; Kahn, A.R.; Shankar, K.; Schneck, M.B.; Weiss, D.J.; Heise, R.L. Porcine Lung-Derived Extracellular Matrix Hydrogel Properties Are Dependent on Pepsin Digestion Time. *Tissue Engineering Part C: Methods* **2020**, *26*, 332-346, doi:10.1089/ten.tec.2020.0042.
23. Gasek, N.; Park, H.E.; Uriarte, J.J.; Uhl, F.E.; Pouliot, R.A.; Riveron, A.; Moss, T.; Phillips, Z.; Louie, J.; Sharma, I.; et al. Development of alginate and gelatin-based pleural and tracheal sealants. *Acta Biomaterialia* **2021**, *131*, 222-235, doi:10.1016/j.actbio.2021.06.048.
24. Zhang, J.; Zheng, W.; Yuan, Z.; Chen, J.; Pei, J.; Xue, P. Pressure Effect on the Rheological Behavior of Highly Filled Solid Propellant During Extrusion Flow. *Polymers* **2025**, *17*, doi:10.3390/polym17152003.
25. Dörr, D.; Standau, T.; Murillo Castellón, S.; Bonten, C.; Altstädt, V. Rheology in the Presence of Carbon Dioxide (CO2) to Study the Melt Behavior of Chemically Modified Polylactide (PLA). *Polymers* **2020**, *12*, doi:10.3390/polym12051108.
26. Harden, F.; Kargar, M.; Handge, U.A.; Mari, M. Influence of Carbon Dioxide on the Glass Transition of Styrenic and Vinyl Pyridine Polymers: Comparison of Calorimetric, Creep, and Rheological Experiments. *Advances in Polymer Technology* **2022**, *2022*, 1-15, doi:10.1155/2022/5602902.
27. Burger, N.A.; Meier, G.; Bouteiller, L.; Loppinet, B.; Vlassopoulos, D. Dynamics and Rheology of Supramolecular Assemblies at Elevated Pressures. *The Journal of Physical Chemistry B* **2022**, *126*, 6713-6724, doi:10.1021/acs.jpcb.2c03295.
28. Ahuja, A.; Lee, R.; Joshi, Y.M. Advances and challenges in the high-pressure rheology of complex fluids. *Advances in Colloid and Interface Science* **2021**, *294*, doi:10.1016/j.cis.2021.102472.
29. Fan, W.; Ji, Y. A novel theoretical strategy for modeling viscosity of polymer melts: Modified friction theory with PC-SAFT. *Fluid Phase Equilibria* **2021**, *550*, doi:10.1016/j.fluid.2021.113262.
30. Azmi, M.A.; Ariff, Z.M.; Shuib, R.K.; Rusli, A.; Ku Ishak, K.M.; Shafiq, M.D.; Abdul Hamid, Z.A.; Zakaria, Z.; Abu Bakar, M.H.; Abdullah, M.K. Effect of Different Pressures on Polymer Flow at a Contraction Path: A Real-Time Numerical Approach. *Makara Journal of Technology* **2024**, *28*, 80-88, doi:10.7454/mst.v28i3.1678.
31. Wang, S.; Sadowski, G.; Ji, Y. Strategy of Coupling Artificial Intelligence with Thermodynamic Mechanism for Predicting Complex Polymer Viscosities. *ACS Sustainable Chemistry & Engineering* **2024**, *12*, 4631-4643, doi:10.1021/acssuschemeng.3c08185.
32. Liao, Y.; Hu, Y.; Tan, Y.; Ikeda, K.; Okabe, R.; Wu, R.; Ozaki, R.; Xu, Q.; Munoz-Espi, R. Measurement Techniques and Methods for the Pressure Coefficient of Viscosity of Polymer Melts. *Advances in Polymer Technology* **2023**, *2023*, 1-15, doi:10.1155/2023/2020247.
33. Chen, J.-Y.; Chen, X.-P.; Huang, M.-S. In-situ measurement of transient polymer melt flow at injection mold gate using pressure sensing and mass conservation principle. *Measurement* **2025**, *256*,

doi:10.1016/j.measurement.2025.118246.
34. Münstedt, H. Influence of hydrostatic pressure on rheological properties of polymer melts—A review. *Journal of Rheology* **2020**, *64*, 751-774, doi:10.1122/1.5142059.
35. Ren, D.; Li, H. A Review of High-Temperature and High-Pressure Experimental Apparatus Capable of Generating Differential Stress. *Frontiers in Earth Science* **2022**, *10*, doi:10.3389/feart.2022.852403.
36. Tianfu, H.; Meinan, L.; Yanbin, C.; Qi, W. Research on the Pressure Control Technology of Ultra-high Temperature & High Pressure Rheometer. *Procedia Engineering* **2014**, *73*, 172-177, doi:10.1016/j.proeng.2014.06.185.
37. Cardinaels, R.; Van Puyvelde, P.; Moldenaers, P. Evaluation and comparison of routes to obtain pressure coefficients from high-pressure capillary rheometry data. *Rheologica Acta* **2006**, *46*, 495-505, doi:10.1007/s00397-006-0148-5.
38. Park, H.E.; Lim, S.T.; Laun, H.M.; Dealy, J.M. Measurement of pressure coefficient of melt viscosity: drag flow versus capillary flow. *Rheologica Acta* **2008**, *47*, 1023-1038, doi:10.1007/s00397-008-0296-x.
39. Wohl, A. Untersuchungen über die Zustandsgleichung. *Zeitschrift für Physikalische Chemie* **1921**, *99U*, 234-241, doi:10.1515/zpch-1921-9918.
40. Tait, P.G. Report on some of the physical properties of fresh water and of sea-water. In *Report on the Scientific Results of the Voyage of H.M.S. Challenger during the Years 1873-1876: Physics and Chemistry, Vol II, Part IV 1888*, Tait, P.G., Ed.; Cambridge University Press: Cambridge, 1900; pp. 1-355.
41. Cutler, W.G.; McMickle, R.H.; Webb, W.; Schiessler, R.W. Study of the Compressions of Several High Molecular Weight Hydrocarbons. *The Journal of Chemical Physics* **1958**, *29*, 727-740, doi:10.1063/1.1744583.
42. Bird, R.B.S., Warren E.; Lightfoot, Edwin N. *Transport Phenomena*, Revised 2nd Edition ed.; John Wiley & Sons: Danvers, MA, 2007.
43. Zoller, P.W., David J. *Standard Pressure-Volume-Temperature Data for Polymers*; Technomic Publishing: Lancaster, Pennsylvania, 1995.
44. Park, H.E.; Lim, S.T.; Smillo, F.; Dealy, J.M.; Robertson, C.G. Wall slip and spurt flow of polybutadiene. *Journal of Rheology* **2008**, *52*, 1201-1239, doi:10.1122/1.2964199.
45. Blanco-Díaz, E.G.; Castrejón-González, E.O.; Rico-Ramírez, V.; Aztatzi-Pluma, D.; Díaz-Ovalle, C.O. Polydispersity influence in rheological behavior of linear chains by molecular dynamics. *Journal of Molecular Liquids* **2018**, *268*, 832-839, doi:10.1016/j.molliq.2018.07.099.
46. Nichetti, D.; Manas-Zloczower, I. Viscosity model for polydisperse polymer melts. *Journal of Rheology* **1998**, *42*, 951-969, doi:10.1122/1.550908.
47. Ye, X.; Sridhar, T. Effects of the Polydispersity on Rheological Properties of Entangled Polystyrene Solutions. *Macromolecules* **2005**, *38*, 3442-3449, doi:10.1021/ma049642n.
48. Park, H.E. Effects of pressure and dissolved carbon dioxide on the rheological properties of molten polymers. McGill University, Montreal, Quebec, Canada, 2005.
49. Arrhenius, S. Über die Reaktionsgeschwindigkeit bei der Inversion von Rohrzucker durch Säuren. *Zeitschrift für Physikalische Chemie* **1889**, *4U*, 226-248, doi:10.1515/zpch-1889-0416.
50. Eyring, H. Viscosity, Plasticity, and Diffusion as Examples of Absolute Reaction Rates. *The Journal of Chemical Physics* **1936**, *4*, 283-291, doi:10.1063/1.1749836.
51. Ewell, R.H.; Eyring, H. Theory of the Viscosity of Liquids as a Function of Temperature and Pressure. *The Journal of Chemical Physics* **1937**, *5*, 726-736, doi:10.1063/1.1750108.
52. Barus, C. Isothermals, isopiestics and isometrics relative to viscosity. *American Journal of Science* **1893**, *s3-45*, 87-96, doi:10.2475/ajs.s3-45.266.87.
53. Batschinski, A.J. Untersuchungen Aber die innere Reibnng der Flüssigkeiten. I. *Zeitschrift für Physikalische Chemie* **1913**, *84U*, 643-706, doi:10.1515/zpch-1913-8442.
54. Williams, M.L.; Landel, R.F.; Ferry, J.D. The Temperature Dependence of Relaxation Mechanisms in Amorphous Polymers and Other Glass-forming Liquids. *Journal of the American Chemical Society* **2002**, *77*, 3701-3707, doi:10.1021/ja01619a008.
55. Park, H.E.; Dealy, J.; Münstedt, H. Influence of long-chain branching on time-pressure and time-temperature shift factors for polystyrene and polyethylene. *Rheologica Acta* **2006**, *46*, 153-159, doi:10.1007/s00397-006-0116-0.
56. Ferry, J.D.; Stratton, R.A. The free volume interpretation of the dependence of viscosities and viscoelastic relaxation times on concentration, pressure, and tensile strain. *Kolloid-Zeitschrift* **1960**, *171*, 107-111, doi:10.1007/bf01520041.
57. Matheson, A.J. Role of Free Volume in the Pressure Dependence of the Viscosity of Liquids. *The Journal of Chemical Physics* **1966**, *44*, 695-699, doi:10.1063/1.1726747.

58. Park, H.E.; Dealy, J.M.; Marchand, G.R.; Wang, J.; Li, S.; Register, R.A. Rheology and Structure of Molten, Olefin Multiblock Copolymers. *Macromolecules* **2010**, *43*, 6789-6799, doi:10.1021/ma1012122.
59. Li, S.W.; Park, H.E.; Dealy, J.M. Evaluation of molecular linear viscoelastic models for polydisperse H polybutadienes. *Journal of Rheology* **2011**, *55*, 1341-1373, doi:10.1122/1.3635384.
60. Kadijk, S.E.; Van Den Brule, B.H.A.A. On the pressure dependency of the viscosity of molten polymers. *Polymer Engineering & Science* **2004**, *34*, 1535-1546, doi:10.1002/pen.760342004.
61. Sedlacek, T.; Lengalova, A.; Zatloukal, M.; Cermak, R.; Saha, P. Pressure and Temperature Dependence of LDPE Viscosity and Free Volume: The Effect of Molecular Structure. *International Polymer Processing* **2006**, *21*, 98-103, doi:10.3139/217.1909.
62. Van Krevelen, D.W.T.N., K. *Properties of Polymers: Their Correlation with Chemical Structure; Their Numerica Estimation and Prediction from Additive Group Contributions*, 4th ed.; Elsevier: Amsterdam, 2009.
63. Cogswell, F.N.; McGowan, J.C. The effects of pressure and temperature upon the viscosities of liquids with special reference to polymeric liquids. *British Polymer Journal* **2007**, *4*, 183-198, doi:10.1002/pi.4980040304.
64. Tschoegl, N.W.; Knauss, W.G.; Emri, I. The Effect of Temperature and Pressure on the Mechanical Properties of Thermo- and/or Piezorheologically Simple Polymeric Materials in Thermodynamic Equilibrium – A Critical Review. *Mechanics of Time-Dependent Materials* **2002**, *6*, 53-99, doi:10.1023/a:1014421519100.
65. Pacheco, J.E.L.; Bavastri, C.A.; Pereira, J.T. Viscoelastic Relaxation Modulus Characterization Using Prony Series. *Latin American Journal of Solids and Structures* **2015**, *12*, 420-445, doi:10.1590/1679-78251412.
66. Li, S.W.; Park, H.E.; Dealy, J.M.; Maric, M.; Lee, H.; Im, K.; Choi, H.; Chang, T.; Rahman, M.S.; Mays, J. Detecting Structural Polydispersity in Branched Polybutadienes. *Macromolecules* **2010**, *44*, 208-214, doi:10.1021/ma101803h.
67. Lee, P.C.; Park, H.E.; Morse, D.C.; Macosko, C.W. Polymer-polymer interfacial slip in multilayered films. *Journal of Rheology* **2009**, *53*, 893-915, doi:10.1122/1.3114370.
68. Park, H.E.L., Patrick C.; Macosko, Christopher W. Polymer-polymer interfacial slip by direct visualization and by stress reduction. *Journal of Rheology* **2010**, *54*, 1207-1218, doi:10.1122/1.3479389.
69. Kim, E.S.; Park, H.E.; Lee, P.C. In situ shrinking fibers enhance strain hardening and foamability of linear polymers. *Polymer* **2018**, *136*, 1-5, doi:10.1016/j.polymer.2017.12.040.
70. Kim, E.S.; Park, H.E.; Lopez-Barron, C.R.; Lee, P.C. Enhanced Foamability with Shrinking Microfibers in Linear Polymer. *Polymers* **2019**, *11*, doi:10.3390/polym11020211.
71. Plazek, D.J. Temperature Dependence of the Viscoelastic Behavior of Polystyrene. *The Journal of Physical Chemistry* **2002**, *69*, 3480-3487, doi:10.1021/j100894a039.
72. Zorn, R.; McKenna, G.B.; Willner, L.; Richter, D. Rheological Investigation of Polybutadienes Having Different Microstructures over a Large Temperature Range. *Macromolecules* **2002**, *28*, 8552-8562, doi:10.1021/ma00129a014.
73. Wood-Adams, P.; Costeux, S. Thermorheological Behavior of Polyethylene: Effects of Microstructure and Long Chain Branching. *Macromolecules* **2001**, *34*, 6281-6290, doi:10.1021/ma0017034.
74. Raju, V.R.; Rachapudy, H.; Graessley, W.W. Properties of amorphous and crystallizable hydrocarbon polymers. IV. Melt rheology of linear and star-branched hydrogenated polybutadiene. *Journal of Polymer Science: Polymer Physics Edition* **2003**, *17*, 1223-1235, doi:10.1002/pol.1979.180170707.
75. Kulisiewicz, L.D., Antonio. High-pressure Rheological Measurement Methods: A Review. *Applied Rheology* **2009**, *20*, 13018.
76. Couch, M.A.; Binding, D.M. High pressure capillary rheometry of polymeric fluids. *Polymer* **2000**, *41*, 6323-6334, doi:10.1016/s0032-3861(99)00865-4.
77. Hubmann, M.; Schuschnigg, S.; Đuretek, I.; Groten, J.; Holzer, C. Enhancing High-Pressure Capillary Rheometer Viscosity Data Calculation with the Propagation of Uncertainties for Subsequent Cross-Williams, Landel, and Ferry (WLF) Parameter Fitting. *Polymers* **2023**, *15*, doi:10.3390/polym15143147.
78. Mackley, M.R.; Marshall, R.T.J.; Smeulders, J.B.A.F. The multipass rheometer. *Journal of Rheology* **1995**, *39*, 1293-1309, doi:10.1122/1.550637.
79. Calvignac, B.; Rodier, E.; Letourneau, J.-J.; Vitoux, P.; Aymonier, C.; Fages, J. Development of an improved falling ball viscometer for high-pressure measurements with supercritical CO2. *The Journal of Supercritical Fluids* **2010**, *55*, 96-106, doi:10.1016/j.supflu.2010.07.012.
80. Mattischek, J.-P.; Sobczak, R. High-pressure cell for measuring the zero-shear viscosity of polymer melts. *Review of Scientific Instruments* **1997**, *68*, 2101-2105, doi:10.1063/1.1148076.
81. Giacomin, A.J.; Samurkas, T.; Dealy, J.M. A novel sliding plate rheometer for molten plastics. *Polymer*

*Engineering & Science* **1989**, *29*, 499-504, doi:10.1002/pen.760290803.
82. Dealy, J.M.; Giacomin, A.J. Sliding plate and sliding cylinder rheometers. In *Rheological Measurement*; 1998; pp. 237-259.
83. Kolitawong, C.; Jeffrey Giacomin, A. Dynamic response of a shear stress transducer in the sliding plate rheometer. *Journal of Non-Newtonian Fluid Mechanics* **2002**, *102*, 71-96, doi:10.1016/s0377-0257(01)00139-2.
84. Koran, F.; Dealy, J.M. A high pressure sliding plate rheometer for polymer melts. *Journal of Rheology* **1999**, *43*, 1279-1290, doi:10.1122/1.551046.
85. Dealy, J.M.R., Daniel J.; Larson, Ronald G. *Structure and Rheology of Molten Polymers; From Structure to Flow Behavoir and Back Again*, 2nd ed.; Hanser: Cincinnati, 2018.
86. Münstedt, H. *Deformation and Flow of Polymeric Materials*; Springer: Heidelberg, 2014.
87. García-Franco, C.A.; Harrington, B.A.; Lohse, D.J. Effect of Short-Chain Branching on the Rheology of Polyolefins. *Macromolecules* **2006**, *39*, 2710-2717, doi:10.1021/ma052581o.
88. Chayrattanaroj, T.; Anantawaraskul, S.; Soares, J.B.P. Using Probability Models to Design the Microstructure of Linear Olefin Block Copolymers. *Macromolecules* **2024**, *57*, 6049-6068, doi:10.1021/acs.macromol.3c02262.
89. Costeux, S.; Anantawaraskul, S.; Wood-Adams, P.M.; Soares, J.B.P. Distribution of the Longest Ethylene Sequence in Ethylene/α-Olefin Copolymers Synthesized with Single-Site-Type Catalysts. *Macromolecular Theory and Simulations* **2002**, *11*, doi:10.1002/1521-3919(20020301)11:3<326::Aid-mats326>3.0.Co;2-z.
90. Wood-Adams, P.M.; Dealy, J.M.; deGroot, A.W.; Redwine, O.D. Effect of Molecular Structure on the Linear Viscoelastic Behavior of Polyethylene. *Macromolecules* **2000**, *33*, 7489-7499, doi:10.1021/ma991533z.
91. Stadler, F.J.; Gabriel, C.; Münstedt, H. Influence of Short-Chain Branching of Polyethylenes on the Temperature Dependence of Rheological Properties in Shear. *Macromolecular Chemistry and Physics* **2007**, *208*, 2449-2454, doi:10.1002/macp.200700267.
92. Rahman, M.S.; Aggarwal, R.; Larson, R.G.; Dealy, J.M.; Mays, J. Synthesis and Dilute Solution Properties of Well-Defined H-Shaped Polybutadienes. *Macromolecules* **2008**, *41*, 8225-8230, doi:10.1021/ma801646u.
93. He, C.; Costeux, S.; Wood-Adams, P.; Dealy, J.M. Molecular structure of high melt strength polypropylene and its application to polymer design. *Polymer* **2003**, *44*, 7181-7188, doi:10.1016/j.polymer.2003.09.009.
94. McKee, M.G.; Unal, S.; Wilkes, G.L.; Long, T.E. Branched polyesters: recent advances in synthesis and performance. *Progress in Polymer Science* **2005**, *30*, 507-539, doi:10.1016/j.progpolymsci.2005.01.009.
95. Torres, E.; Li, S.-W.; Costeux, S.; Dealy, J.M. Branching structure and strain hardening of branched metallocene polyethylenes. *Journal of Rheology* **2015**, *59*, 1151-1172, doi:10.1122/1.4927919.
96. Wood-Adams, P.M.; Dealy, J.M. Using Rheological Data To Determine the Branching Level in Metallocene Polyethylenes. *Macromolecules* **2000**, *33*, 7481-7488, doi:10.1021/ma991534r.
97. Shroff, R.N.; Mavridis, H. Long-Chain-Branching Index for Essentially Linear Polyethylenes. *Macromolecules* **1999**, *32*, 8454-8464, doi:10.1021/ma9909354.
98. Graessley, W.W. Effect of long branches on the temperature dependence of viscoelastic properties in polymer melts. *Macromolecules* **2002**, *15*, 1164-1167, doi:10.1021/ma00232a040.
99. Malpass, D.B. *Introduction to Industrial Polyethylene: Properties, Catalysts, Processes*; John Wiley & Sons: Danvers, MA, 2010.
100. Kim, S. A criterion for gross melt fracture of polyolefins and its relationship with molecular structure McGill University, 2000.
101. Park, H.E. Effect of pressure on the rheological properties of three polyethylenes. McGill University, Montreal, 2001.
102. Kim, S.; Dealy, J.M. Gross melt fracture of polyethylene. I: A criterion based on tensile stress. *Polymer Engineering & Science* **2002**, *42*, 482-494, doi:10.1002/pen.10965.
103. Wood-Adams, P.M. Determination of molecular weight distribution from rheological measurements. McGill University, Montreal, Canada, 1995.
104. Zoller, P.; Bolli, P.; Pahud, V.; Ackermann, H. Apparatus for measuring pressure–volume–temperature relationships of polymers to 350 °C and 2200 kg/cm2. *Review of Scientific Instruments* **1976**, *47*, 948-952, doi:10.1063/1.1134779.
105. Jeong, S.; Kim, J.M.; Cho, S.; Baig, C. Effect of short-chain branching on interfacial polymer structure and dynamics under shear flow. *Soft Matter* **2017**, *13*, 8644-8650, doi:10.1039/c7sm01644a.